\documentclass[aps,prx,twocolumn,english,superscriptaddress,floatfix,longbibliography,10pt]{revtex4-2}
\usepackage[unicode]{hyperref}
\usepackage{float}
\hypersetup{unicode=true,colorlinks=true,linkcolor=blue, citecolor=blue,urlcolor=blue}

\usepackage{booktabs}
\usepackage{multirow}
\usepackage{graphicx}
\usepackage{dcolumn}
\usepackage{bm}
\usepackage{soul}

\usepackage{mathrsfs}
\usepackage{amsfonts}
\usepackage{amsmath}
\usepackage{amsthm}
\usepackage{amssymb}
\usepackage{physics}
\usepackage{dsfont}
\usepackage{esint}
\usepackage[linesnumbered,ruled,vlined]{algorithm2e}

\usepackage[dvipsnames, svgnames, x11names]{xcolor}
\newcommand{\catp}{\mathcal{C}_\alpha^+} 
\newcommand{\catm}{\mathcal{C}_\alpha^-} 
\newcommand{\Pcat}{P_{\rm{cat}}}
\newcommand{\Zcat}{Z_{\rm{cat}}}
\newcommand{\Xcat}{X_{\rm{cat}}}
\newcommand{\Ycat}{Y_{\rm{cat}}}
\newcommand{\acat}{a_{\rm{cat}}}

\begin{document}

\title{Fault-tolerant quantum computing with a microwave Cat Bus}

\author{Yanyan Chen}
\thanks{These authors contributed equally to this work.}
\affiliation{State Key Laboratory of Surface Physics, Institute of Nanoelectronics and Quantum Computing, and Department of Physics, Fudan University, Shanghai 200433, China}
\affiliation{Shanghai Qi Zhi Institute, AI Tower, Xuhui District, Shanghai 200232, China} 

\author{Xinyang Yu}
\thanks{These authors contributed equally to this work.}
\affiliation{State Key Laboratory of Surface Physics, Institute of Nanoelectronics and Quantum Computing, and Department of Physics, Fudan University, Shanghai 200433, China}
\affiliation{Shanghai Qi Zhi Institute, AI Tower, Xuhui District, Shanghai 200232, China}

\author{Yueyang Min}
\affiliation{State Key Laboratory of Surface Physics, Institute of Nanoelectronics and Quantum Computing, and Department of Physics, Fudan University, Shanghai 200433, China}
\affiliation{Shanghai Qi Zhi Institute, AI Tower, Xuhui District, Shanghai 200232, China} 

\author{Zhihao Zhang}
\affiliation{State Key Laboratory of Surface Physics, Institute of Nanoelectronics and Quantum Computing, and Department of Physics, Fudan University, Shanghai 200433, China}
\affiliation{Shanghai Qi Zhi Institute, AI Tower, Xuhui District, Shanghai 200232, China}

\author{Shuaifan Cao}
\affiliation{State Key Laboratory of Surface Physics, Institute of Nanoelectronics and Quantum Computing, and Department of Physics, Fudan University, Shanghai 200433, China}
\affiliation{Shanghai Qi Zhi Institute, AI Tower, Xuhui District, Shanghai 200232, China}

\author{Xiaopeng Li}
\affiliation{State Key Laboratory of Surface Physics, Institute of Nanoelectronics and Quantum Computing, and Department of Physics, Fudan University, Shanghai 200433, China}
\affiliation{Shanghai Qi Zhi Institute, AI Tower, Xuhui District, Shanghai 200232, China}
\affiliation{Shanghai Buchou Quantum Technology Co., Ltd}
\affiliation{Hefei National Laboratory, Hefei 230088, China}
\email{xiaopeng\underline{ }li@fudan.edu.cn}

\date{\today}

\begin{abstract}
The scalability of fault-tolerant neutral-atom quantum computers is constrained by the latency of shuttling with optical tweezers, imposing a stringent trade-off between qubit overhead and circuit depth in quantum algorithm compilation. Here we propose a hardware-efficient, shuttling-free architecture that achieves all-to-all connectivity. Remote Rydberg atoms are resonantly entangled through a microwave ``Cat Bus''---a cavity mode autonomously stabilized in a bosonic cat state. The Cat Bus natively supports the highly parallelized execution of one-to-many $\mathrm{CZ}^n$ gates with exponentially suppressed crosstalk. We derive the resulting cat--atom error channel from the underlying interactions and physical constraints. For fault-tolerant operation, we develop a hardware-aware scheduling scheme that exploits the native cat--atom $\mathrm{CZ}^{n}$ gate to construct a syndrome-extraction circuit with minimum depth. We benchmark the architecture using hypergraph-product (HGP) codes and estimate a 180-fold reduction in syndrome-extraction cycle time at $N=10^5$ data qubits compared with an atom-rearrangement-based architecture. Under matched two-qubit depolarizing noise, the corresponding error threshold increases from $0.55\%$ to $0.72\%$. Under the hardware-derived error model, we obtain a threshold of $0.80\%$, corresponding to a threshold cooperativity of $C_{\mathrm{th}}=7.8 \times 10^4$, compatible with experimentally accessible parameters for Rydberg-coupled microwave-cavity systems. By avoiding atom transport, the Cat Bus provides a route towards high-speed, fault-tolerant neutral-atom quantum computation.
\end{abstract}

\maketitle

\section{Introduction}
Quantum low-density parity-check (qLDPC) codes offer a highly resource-efficient pathway for scaling fault-tolerant quantum processors~\cite{Breuckmann2021LDPC,Bravyi2024FaultTolerantMemory,Xu2024ConstantOverhead}. With constant encoding rate~\cite{Panteleev2021GoodLDPC} and linear scaling distance~\cite{Panteleev2022GoodLDPC}, good qLDPC codes reduce the overhead of practically relevant algorithms---such as breaking RSA-2048---down to the $10^4\sim10^5$ physical qubit regime~\cite{Iceberg2026Pinnacle,madelyn2026shor}. To physically realize this theoretical advantage, the underlying hardware must support non-local qubit connectivity~\cite{Bravyi2010Tradeoff,Baspin2022Nonlocal}. The neutral-atom platform has emerged as a leading contender to meet this demand,  utilizing dynamic optical tweezers to physically establish long-range interactions~\cite{Bluvstein2024LogicalProcessor,Bluvstein2025FT, Review2026}.  However, syndrome extractions in this architecture are constrained by the latency of mechanical atom transport~\cite{Saffman2025Review,Xu2024ConstantOverhead}. Despite engineering optimizations~\cite{pagano2024optimal, Hwang2025Tweezer}, the kinematics of accelerating atoms impose a fundamental limit on error correction cycle times. This clock-speed bottleneck forces a severe space-time tradeoff: maintaining viable algorithmic runtime necessitates massive parallelization, at the cost of inflating the physical qubit count~\cite{Iceberg2026Pinnacle, madelyn2026shor,Zhou2025Resource}. To bypass these kinematic constraints, an architectural shift is required: transitioning from mechanical rearrangement to high-speed interconnects without physical transport.

Photons provide an ideal medium for such interconnects, while completely bypassing the mechanical noise of physical atom transport~\cite{Ramette2024Interconnect,Sinclair2025Interconnect}. Pioneering experiments in cavity quantum electrodynamics (cQED) have demonstrated that light fields can mediate long-range interactions, effectively transforming static arrays into highly connected quantum processors~\cite{Welte2018Mediated,Monika2021Programmable,Wu2026LongRange,Ye2023Universal}. Traditional light-mediated schemes, however, primarily rely on standard Fock states~\cite{Duan2005Robust,Ramette2022AnyToAny,Monica2025Cavity} or unprotected coherent states~\cite{Sorensen2003Cavity,Jandura2024Cavity}. As a consequence, these approaches remain sensitive to photon loss and lack hardware-level error resilience required for scalable fault tolerance. 

To overcome these limitations while preserving the advantages of photonic interconnects, we introduce a hardware architecture that utilizes cat-code-encoded superconducting microwave cavities~\cite{Mirrahimi2014Cat, Ofek2016Cat, Grimm2020Stabilization, Christopher2022FTQC} to mediate entanglement between remote Rydberg atoms, an approach we dub the ``Cat Bus". 
The large dipole moments of Rydberg states~\cite{saffman2010} enable strong cavity coupling and therefore fast gate times.
Furthermore, cat encoding provides exponentially suppressed bit flips~\cite{Guillaud2019RepetitionCat,Puri2020Bias, Putterman2025BosonicConcatenation}, and the leading errors of cat-atom gates are dominated by cat $Z$ errors and atomic leakage. Crucially, a single cavity mode enables the execution of a one-to-many $\mathrm{CZ}^n$ gate with exponentially suppressed crosstalk in constant time, thus mediating highly parallelized quantum gates among atoms. 

\begin{figure*}[htbp]
    \centering
    \includegraphics[width=\linewidth]{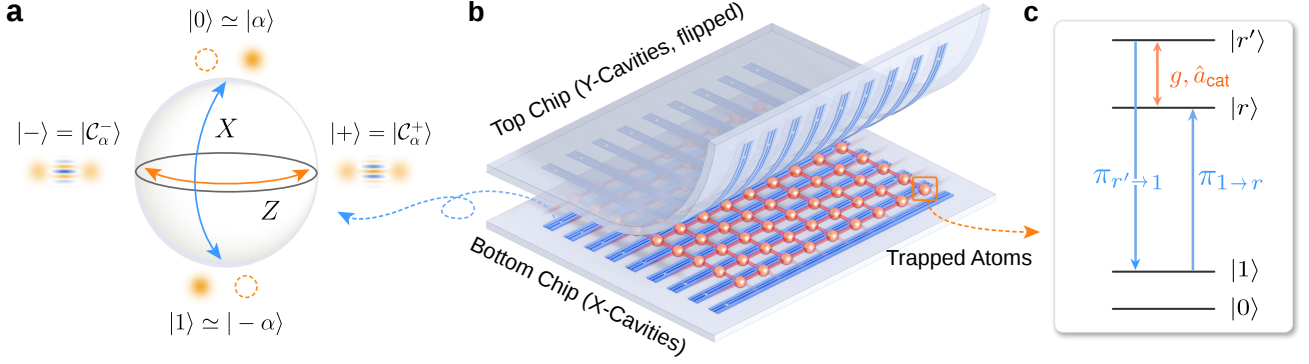}
    \caption{\textbf{A shuttling-free neutral-atom quantum processor mediated by a microwave Cat Bus.} \textbf{a}, Bloch sphere representation of the stabilized cat-state manifold $\mathcal{C} = \text{span}\{\ket{\alpha}, \ket{-\alpha}\}$, where logical Pauli operators are defined in the orthogonal $X$-basis $\ket{\mathcal{C}_\alpha^\pm} \propto (\ket{\alpha} \pm \ket{-\alpha})$.
    \textbf{b}, Crossed-chip hardware configuration. A static 2D optical lattice traps neutral atoms between two chips containing orthogonal arrays of microwave resonators (top Y-cavities and bottom X-cavities). These resonators function as `Cat Buses', mediating long-range entanglement and establishing a highly connected, shuttling-free processor geometry.
    \textbf{c}, Atomic energy level structure. Quantum information is stored in the computational subspace spanned by its hyperfine states $\ket{0}$ and $\ket{1}$. The Rydberg transition $\ket{r} \leftrightarrow \ket{r'}$ is resonant with and strongly coupled to its corresponding row or column cat-qubit mode $\hat{a}_{\text{cat}}$.}
    \label{fig:hardware}
\end{figure*}

To exploit the native multi-target $\mathrm{CZ}^{n}$ interactions for error correction, we develop a hardware-aware scheduling scheme that generalizes conventional edge coloring from pairwise matchings to simultaneously executable one-to-many gates. Within this formulation, minimizing the number of scheduling steps reduces to finding a minimum vertex cover of the underlying Tanner graph. We apply this general scheme to hypergraph-product (HGP) codes, a representative qLDPC family whose product structure allows these gates to be mapped naturally onto parallel row- and column-bus operations.
With $N=10^{5}$ data qubits, the resulting architecture reduces the estimated syndrome-extraction cycle time by a factor of $180$ relative to atom rearrangement. 
Under the same two-qubit depolarizing noise model, the circuit-level threshold increases from 0.55\% for the rearrangement architecture to 0.72\% for the Cat Bus. Under the hardware-derived model of cat dephasing and atomic leakage, it further reaches 0.80\%, corresponding to a threshold cooperativity $C_{\mathrm{th}}=7.8\times10^{4}$, which is experimentally accessible in Rydberg-coupled microwave-cavity systems. These results reveal the system-level advantages, which originate from the inherent parallelism of the cavity-mediated \(\mathrm{CZ}^{n}\) gate and the optimized syndrome-extraction scheduling. More broadly, 
it demonstrates how co-design across the hardware and error-correction layers translates native multi-target interactions and structured noise into concrete improvements in error-correction speed and fault tolerance, guiding the design for scalable fault-tolerant quantum architectures.

\section{A Shuttling-Free Architecture with All-to-all Connectivity}

We propose a hybrid architecture coupling neutral atoms with superconducting microwave cavities in a crossed-chip geometry. As illustrated in Fig.~\ref{fig:hardware}, the top flipped chip and the bottom chip contain arrays of equally spaced coplanar waveguide resonators (CPWR)~\cite{Gppl2008Coplanar,Blais2020CQED} oriented along the $\mathrm{Y}$ and $\mathrm{X}$ axes, respectively. Neutral atoms are trapped by a static two-dimensional optical lattice, sandwiched between the two chips. Specifically, each atom levitates at the intersection of resonator modes of a $\mathrm{Y}$-cavity and an $\mathrm{X}$-cavity~\cite{Wilde2025Hybrid}. 

Each resonator is encoded and stabilized as a bosonic cat qubit via strong two-photon dissipation with rate $\kappa_2$~\cite{Guillaud2019RepetitionCat,Christopher2022FTQC,Putterman2025BosonicConcatenation}. 
This dissipation restricts the resonator to a two-dimensional cat manifold $\mathcal{C} = \mathrm{span}\{|\alpha\rangle, |-\alpha\rangle\}$.
In the large-amplitude limit ($|\alpha|^2 \gg 1$), the overlap $u \equiv \langle -\alpha|\alpha\rangle = e^{-2|\alpha|^2}$ is exponentially suppressed, rendering the coherent states $|\pm \alpha\rangle$ asymptotically orthogonal. Consequently, as shown in Fig.~\ref{fig:hardware}\textbf{a}, these states $\ket{\pm \alpha}$ serve as the computational basis states $\ket{0}_{\mathrm{cat}}$ and $\ket{1}_{\mathrm{cat}}$ for the individual bosonic cat qubit.

Individual atoms are selected using cross-point addressing with orthogonal laser beams in the $\mathrm{X}$ and $\mathrm{Y}$ directions~\cite{Weiss2015Addressing,Weiss2016Addressing}. 
Through energy-level dressing, only the atom at the beam intersection is shifted into resonance with a global Rydberg excitation pulse. 
Multiple cross-points can be generated simultaneously, allowing the $|1\rangle$ component of each addressed atom to be transferred to the Rydberg manifold, where the atoms resonantly couple to the corresponding row or column cavity mode, as illustrated in Fig.~\ref{fig:hardware}\textbf{c}. Each shared cavity mode can thus mediate long-range interactions among the selected atoms, and we refer to it as a ``Cat Bus''. For two atoms that do not share a bus, a third atom at the crossing of their row and column can mediate a constant-depth entangling gate (Supplementary Information). The hybrid architecture therefore transforms the static atom array into a shuttling-free quantum processor with all-to-all connectivity. 

\begin{figure}[htbp]
    \centering
    \includegraphics[width=1\linewidth]{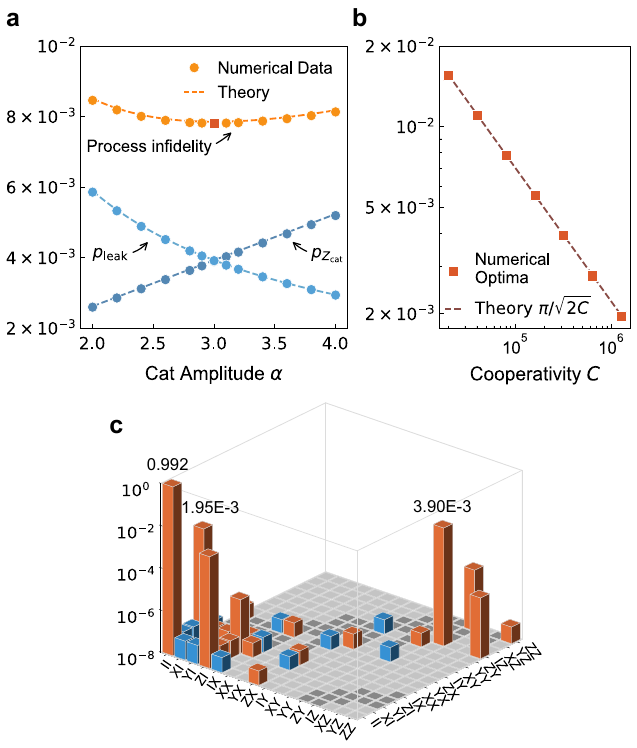}
    \caption{\textbf{Error characterization of the cat-atom $\mathrm{CZ}^1$ gate.} \textbf{a}, Process infidelity as a function of the cat amplitude $\alpha$ at a fixed ratio $\gamma/\kappa_1 = 18$. The trade-off between the two dominant error contributions---photon loss ($\propto \alpha \kappa_1$) and Rydberg decay ($\propto \gamma/\alpha$) gives rise to the optimal performance at $\alpha_{\rm{opt}} = \sqrt{\gamma/(2\kappa_1)} = 3$.
    \textbf{b}, Optimized process infidelity versus cooperativity $C = g^2/(\kappa_1 \gamma)$, obtained by minimizing over $\alpha$ at fixed ratio in \textbf{a}. The data follow the scaling $\pi/\sqrt{2C}$ (dashed line).
    \textbf{c}, $\chi$-matrix tomography in the computational subspace spanned by $\{\ket{0},\ket{1}\}\otimes\{\ket{\pm\alpha}\}$. The tomography is evaluated at the parameters of $C = 8 \times 10^4$, $|\alpha|^2 = 9$, $\kappa_1 /g= 8.33 \times 10^{-4}$, and $\gamma/g = 1.50\times10^{-2}$ (corresponding to the third data point in panel \textbf{b}). Orange and blue bars denote positive and negative elements, respectively. The dominant $(IZ,IZ)$, $(ZI,II)$ and $(II,ZI)$ components match $p_{Z_{\mathrm{cat}}}=3.91\times10^{-3}$ and $p_{\mathrm{leak}}/2=1.95\times10^{-3}$ in Eq.~\eqref{eq:CZn_channel}. The resulting process infidelity is $\epsilon_1 = 7.81\times 10^{-3}$, consistent with the theoretical optimum $\epsilon_{\min}=7.85\times 10^{-3}$.
    }
    \label{fig:CZ}
\end{figure}

\section{Native one-to-many cat–atom \texorpdfstring{\(\mathrm{CZ}^{n}\)}{CZn} gates}
\label{sec: III CZ channel}

The fundamental building block of our architecture is a high-speed, multi-qubit entangling operation between a cat qubit and atoms. 
We consider $n$ neutral atoms collectively coupled to a single resonator mode $\hat{a}_{\mathrm{cat}}$, which is tuned to resonance with the Rydberg transition $|r\rangle \leftrightarrow |r'\rangle$. In the strong two-photon dissipative regime ($\kappa_2 \gg g$), the cavity dynamics are autonomously restricted to the stabilized cat manifold $\mathcal{C}$~\cite{Guillaud2019RepetitionCat,Christopher2022FTQC,Putterman2025BosonicConcatenation}.
This constraint is formally described by the projector $P_\mathrm{cat} = \ket{\mathcal{C}_\alpha^+} \bra{\mathcal{C}_\alpha^+} + \ket{\mathcal{C}_\alpha^-} \bra{\mathcal{C}_\alpha^-}$. 
Within the cat manifold, the cavity annihilation operator takes the form $\hat{a}_\mathrm{cat} := P_\mathrm{cat} \hat{a} P_\mathrm{cat}  = \alpha Z_{\mathrm{cat}} + \mathcal{O}(u)$ where $\hat Z_{\mathrm{cat}}$ is the logical Pauli-$Z$ operator in the
coherent-state basis $\{\ket{\alpha},\ket{-\alpha}\}$~\cite{Puri2020Bias}. Consequently, $[\hat a_{\mathrm{cat}},\hat a_{\mathrm{cat}}^\dagger] =\mathcal{O}(u)$ is exponentially suppressed. 

To leading order in $u$, the projected multi-atom Jaynes--Cummings interaction reduces to an atom-cavity coupling proportional to $g\alpha \sum_{j=1}^n\left(\sigma_x^{(j)} \otimes {Z}_{\rm{cat}}\right)$, where $\hat\sigma_x^{(j)}$ is the Pauli-$X$ operator of the $j$-th atom in the Rydberg basis $\{|r\rangle, |r'\rangle\}$. 
Because the constituent terms commute in the large-\(\alpha\) limit, the evolution factorizes as
\begin{equation}\label{eq:factorized_interaction}
    \hat{U}(t) = \prod_{j=1}^n \exp \left(-i g \alpha t \, \sigma_x^{(j)} \otimes {Z}_{\rm{cat}}\right).
\end{equation}
Crucially, this factorization implies that the cat acts as a shared bus, simultaneously implementing the same conditional rotation on the $n$ coupled atoms. At finite \(\alpha\), residual non-commutativity gives rise to a crosstalk infidelity of order $\mathcal{O}(u^2)$, which is negligible for the parameters considered (Methods). By contrast, for a conventional cavity, the non-vanishing commutator \([\hat a,\hat a^\dagger]=1\) renders the resonant exchange terms associated with different atoms non-commuting. The resulting evolution is therefore collective rather than a product of independent parallel cat-atom gates~\cite{Wu2026LongRange,Monica2025Cavity}. Therefore, cat encoding is part of the parallel entangling mechanism itself.

Building on this parallelized interaction, we realize a one-to-many cat-atom gate, denoted $\mathrm{CZ}^n$,  whose ideal logical action is equivalent to that of $\prod_{j=1}^{n}\mathrm{CZ}_{\mathrm{cat},j}$. 
This native gate is implemented via a high-speed, three-step sequence: (i) a local $\pi$-pulse of duration $T_{\pi}=\pi/\Omega$ maps the $\ket{1}$ component to the Rydberg state $\ket{r}$ for all addressed atoms; (ii) the atom--bus interaction is applied for a duration $T_{\mathrm{int}} = \pi/(2g\alpha)$, corresponding to $\hat U(T_{\mathrm{int}})$; and (iii) a return $\pi$-pulse of duration $T_{\pi}$ from the Rydberg state $\ket{r^\prime}$ to $\ket{1}$, followed by local phase corrections $S=\mathrm{diag}\{1,i\}$. Because all addressed atoms interact with the shared cat mode simultaneously, the total gate time $T_{\mathrm{gate}} =  T_{\mathrm{int}} + 2T_{\pi}$ is independent of the number of targets $n$ at fixed $\alpha$. Assuming state-of-the-art parameters $g=2\pi\times1~\mathrm{MHz}$~\cite{Wilde2025Hybrid}, $\Omega=2\pi \times 17~\mathrm{MHz}$~\cite{Evered2023Parallel,Evered2026HighFidelity} and $\alpha=3$, the total gate time is $T_{\mathrm{gate}} \approx 142~\mathrm{ns}$. 

\begin{figure*}[htbp]
    \centering
    \includegraphics[width=\linewidth]{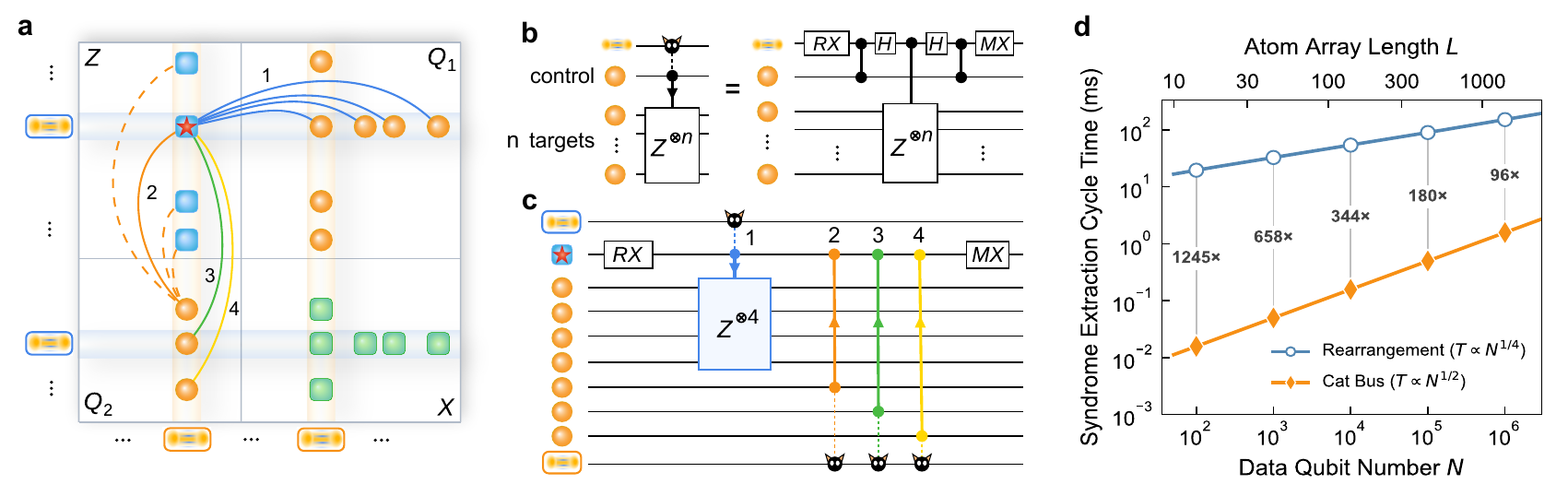}
    \caption{\textbf{Cat-bus-enabled syndrome extraction and cycle-time scaling for hypergraph product (HGP) codes.} \textbf{a}, Physical layout of a hypergraph-product (HGP) code constructed from two identical $(3,4)$-biregular classical Tanner graphs. Orange sites denote data qubits in the $Q_1$ and $Q_2$ blocks, whereas blue and green sites denote $Z$- and $X$-check ancillas, respectively. The marked weight-7 $Z$ check couples to four data qubits in $Q_1$ and three in $Q_2$, and the edge labels 1--4 indicate the four interaction layers shown in \textbf{c}. \textbf{b}, Cat-mediated atom--atom $\mathrm{CZ}^{n}$ gate (left) and its decomposition into the native cat--atom entangling operations of our architecture (right). By design, the cat qubit disentangles from the data atoms in the absence of noise, deterministically yielding a zero outcome upon $X$-basis measurement ($\mathrm{MX}$). When the circuit is noisy, the cat ancilla can function as a built-in error flag.
    \textbf{c}, The syndrome extraction circuit for the marked $Z$-check, decomposed into four scheduling steps induced by the minimum-vertex-cover-based star scheduling. In step 1, the marked $Z$-check acts as the common control to its $Q_1$ neighbors through a horizontal cat bus. In steps 2--4, each of its three $Q_2$ neighbors acts in turn as the control of a column-parallel $\mathrm{CZ}^{4}$ operation targeting four $Z$-check ancillas, including the marked $Z$-check.
    \textbf{d}, Syndrome extraction cycle time $T$ versus data qubit number $N$ and atom array length $L$. 
    The cat-bus estimate (orange) scales as $N^{1/2}$, whereas the atom-rearrangement estimate (blue) scales as $N^{1/4}$. Owing to its substantially smaller prefactor, the Cat Bus remains faster throughout the range shown and yields a 180-fold reduction in cycle time with $N=10^{5}$.
    }
    \label{fig:syndrome_extraction_circuit}
\end{figure*}

The leading-order dissipative error channel of the native $\mathrm{CZ}^{n}$ gate, including Rydberg-state decay (at rate $\gamma$) and cavity single-photon loss (at rate $\kappa_1$), is given by: 
\begin{equation}
    \mathcal E_{\mathrm{CZ}^{n}}=
    \mathcal{Z}_{\mathrm{cat}}(p_{Z_{\mathrm{cat}}})
    \circ
    \prod_{j=1}^{n}
    \mathcal E^{(j)}_{\mathrm{atom}}(p_{\mathrm{leak}}),
    \label{eq:CZn_channel}
\end{equation}
Here, $\mathcal Z_{\mathrm{cat}}$ denotes the cat phase-flip channel, whereas $\mathcal E_{\mathrm{atom}}^{(j)}$ denotes the rydberg leakage channel of atom \(j\) (Methods). Notably, the native \(\mathrm{CZ}^{n}\) gate and \(n\) independently implemented pairwise
\(\mathrm{CZ}\) gates have fundamentally different error channels. The \(X\)-type errors are exponentially suppressed, scaling as \(\mathcal{O}(u)\), consistent with the process tomography in Fig.~\ref{fig:CZ}. 
Treating the local pulses as ideal, the leading-order total process infidelity of the native \(\mathrm{CZ}^{n}\) gate is
\begin{equation}
    \epsilon_n(\alpha) = p_{Z_{\mathrm{cat}}} + n p_{\mathrm{leak}} 
                        = \frac{\pi}{2g} \left( \kappa_1\alpha + \frac{n\gamma}{2\alpha} \right) .
\end{equation}
The two terms arise from the dominant error processes, which exhibit opposite dependences on the cat amplitude. Cavity single-photon loss induces a \(Z_{\mathrm{cat}}\) error with probability $p_{Z_{\mathrm{cat}}}=\kappa_1\alpha^2T_{\mathrm{int}}$, whereas Rydberg decay produces leakage to each addressed atom with probability $p_{\mathrm{leak}}=\gamma T_{\mathrm{int}}/2$ (Methods). 

As \(\alpha\) increases, the photon-loss contribution increases, whereas the shorter interaction time reduces atomic leakage. (Fig.~\ref{fig:CZ}\textbf{a}). Minimizing \(\epsilon_n(\alpha)\) with respect to \(\alpha\) gives
\begin{equation}
        \alpha_{\mathrm{opt}}^{(n)}
    =\sqrt{\frac{n\gamma}{2\kappa_1}},
    \qquad
    \epsilon_{n,\min}
    =\pi\sqrt{\frac{n}{2C}},
\end{equation}
where \(C=g^2/(\kappa_1\gamma)\) is the cooperativity. 
For \(n=1\) and
\(\gamma/\kappa_1=18\), the optimal amplitude is
\(\alpha_{\mathrm{opt}}=3\), matching the minimum in
Fig.~\ref{fig:CZ}\textbf{a}. The corresponding minimum infidelity,
\(\epsilon_{1,\min}=\pi/\sqrt{2C}\), follows the numerical scaling in
Fig.~\ref{fig:CZ}\textbf{b}. At \(C=8\times10^4\), the predicted value
\(7.85\times10^{-3}\) agrees with the process-tomography result
\(7.81\times10^{-3}\) in Fig.~\ref{fig:CZ}\textbf{c}. 

\section{Scalable Fault-Tolerance with HGP Codes}
We benchmark the shuttling-free architecture using hypergraph-product (HGP) codes, a constant-rate family of quantum low-density parity-check (qLDPC) codes~\cite{Tillich2009HGP,Breuckmann2021LDPC}. Without loss of generality, we consider a family constructed from two identical $(3,4)$-biregular classical Tanner graphs, with variable and check degrees $\Delta_{\mathrm V}=3$ and $\Delta_{\mathrm C}=4$, respectively. As illustrated in
Fig.~\ref{fig:syndrome_extraction_circuit}\textbf{a}, the atom array is partitioned into data and check qubits, and each stabilizer check is of weight 7.

The product structure of the HGP code maps naturally onto the crossed row- and column-bus geometry. For either an $X$- or a $Z$-check measurement, the required check--data interactions decompose into horizontal and vertical sectors. Each sector comprises repeated copies of one component classical Tanner graph and maps onto the corresponding row or column Cat Buses. Because the two sectors must be executed sequentially, their contributions to the syndrome-extraction depth are additive and can be minimized independently. For the identical
component graphs considered here, both optimizations reduce to the same one-dimensional classical scheduling problem.

As shown in Fig.~\ref{fig:syndrome_extraction_circuit}\textbf{b}, we combine the native cat--atom interactions to realize a cat-mediated atom--atom $\mathrm{CZ}^{n}$ gate. This multi-target gate provides the elementary operation for the resulting one-dimensional scheduling problem. In the underlying classical Tanner graph, each such operation implements a star, namely a set of edges incident on a common center. Each selected star defines one step of the one-dimensional scheduling. Within this hardware-constrained formulation, minimizing the number of scheduling steps is equivalent to finding a minimum vertex cover (MVC), the smallest set of vertices incident on every edge of the Tanner graph. 

The MVC-optimized one-dimensional star-based scheduling is then lifted onto the HGP qubit array to yield the two-dimensional product scheduling (Methods). By replacing matching-based pairwise execution with native star-wise parallelism, this construction aligns HGP syndrome extraction with the Cat-Bus interaction structure and exemplifies co-design between hardware and quantum error correction.

Applying this star-based scheduling to the marked weight-7 $Z$ check in
Fig.~\ref{fig:syndrome_extraction_circuit}\textbf{a} yields the four
scheduling steps shown in
Fig.~\ref{fig:syndrome_extraction_circuit}\textbf{c}. In Step~1, the marked $Z$-check ancilla acts as the control of a cat-mediated atom--atom $\mathrm{CZ}^{4}$ gate on its four neighboring data qubits in the $Q_1$ block, which share the same horizontal Cat Bus. The corresponding $\mathrm{CZ}^{4}$ operations can be executed simultaneously across independent rows. Completing all such horizontal interactions between the $Z$-check ancillas and the $Q_1$ data block requires $n_{\mathrm C}$ scheduling steps, where $n_{\mathrm C}$ is the number of check nodes in the underlying $(3,4)$-biregular classical code.

For the remaining three data-qubit neighbors in the $Q_2$ block, the protocol switches to column-parallelized operations (Steps~2--4). In each step, a $Q_2$ data qubit acts as the control of a cat-mediated atom--atom $\mathrm{CZ}^{4}$ gate targeting four $Z$-check ancillas through a vertical Cat Bus. Note that the marked $Z$ check is one of the four targets. Completing all vertical interactions between the $Q_2$ data block and the $Z$-check ancillas requires a further $n_{\mathrm C}$ scheduling steps. Thus, each $Z$- or $X$-check round is completed in $2n_{\mathrm C}$ scheduling steps. Because the two check rounds are performed sequentially, a complete syndrome-extraction cycle requires $4n_{\mathrm C}$ scheduling steps.

\subsection{Cycle-Time Performance and Scaling}

The syndrome-extraction cycle time, $T_{\mathrm{cycle}}$, determines the error-correction clock rate of the fault-tolerant processor. As established above, a complete cycle contains $4n_{\mathrm C}$ sequential scheduling
steps, while the cat-mediated operations assigned to distinct row or column Cat Buses within each step are executed in parallel. Each scheduling step requires a cat-mediated atom-atom $\mathrm{CZ}^n$ block in Fig.~\ref{fig:syndrome_extraction_circuit}\textbf{b}. Including cat-qubit state preparation and measurement (SPAM), the total cycle time is
\begin{equation}
    T_{\mathrm{cycle}}^{\mathrm{cat}}
    =
    4n_{\mathrm{C}}
    \left(3T_{\mathrm{gate}}+T_{\mathrm{SPAM}}\right),
    \label{eq:cat_cycle_time}
\end{equation}
where
$n_{\mathrm C}=L\Delta_{\mathrm V}/ (\Delta_{\mathrm C}+\Delta_{\mathrm V})$ for a biregular classical Tanner graph, and $\left( 3T_{\mathrm{gate}}+T_{\mathrm{SPAM}} \right)$ is the gate time of the cat-mediated atom-atom $\mathrm{CZ}^n$ block, with $T_{\mathrm{SPAM}}=T_{\mathrm{MX}}+T_{\mathrm{RX}}\approx 229\,\mathrm{ns}$ (Supplementary Information). 
The durations of the two cat-qubit Hadamard gates are neglected because of their much shorter timescales.

For fixed Tanner-graph degrees, the number of data qubits scales as $N\propto L^2$. Equation~\eqref{eq:cat_cycle_time} therefore gives $T_{\mathrm{cycle}}^{\mathrm{cat}}\propto L\propto N^{1/2}$. Figure~\ref{fig:syndrome_extraction_circuit}\textbf{d} compares this result with the atom-rearrangement estimate of Xu \emph{et al.}~\cite{Xu2024ConstantOverhead}, which scales as $N^{1/4}$ (Supplemtary Information). The nanosecond-scale entangling time $T_{\mathrm{gate}}\approx 142\,\mathrm{ns}$, together with the comparably short SPAM time, yields a substantially smaller cycle-time prefactor than atom rearrangement. Consequently, the Cat Bus reduces the syndrome-extraction cycle time by factors ranging from approximately $10^{2}$ to $10^{3}$ over the system sizes shown. With $N=10^{5}$ data qubits, our architecture achieves a 180-fold reduction in the estimated syndrome-extraction cycle time.

\subsection{Fault-Tolerant Thresholds and Sub-threshold Scaling}
\begin{figure}[htbp]
    \centering
    \includegraphics[width=\linewidth]{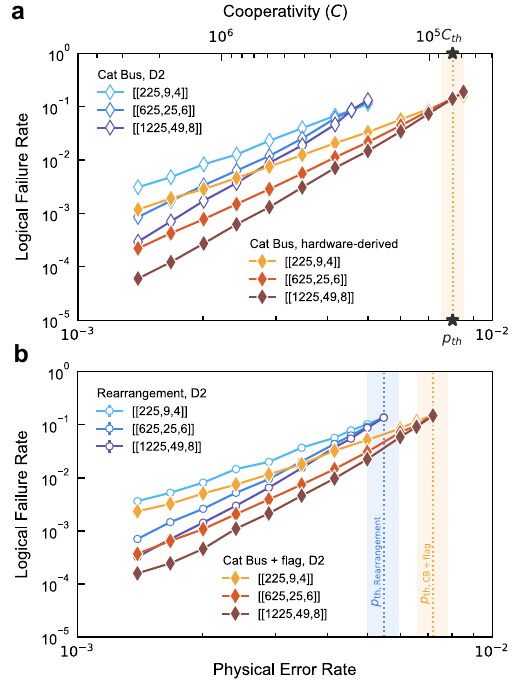}
    \caption{\textbf{Circuit-level logical performance of the Cat Bus architecture.} \textbf{a,} Logical failure rate per syndrome-extraction cycle for the same Cat Bus circuits under the two-qubit depolarizing (D2) model and the hardware-derived (HD) Cat Bus error model. The upper axis converts the physical error rate to cooperativity using $p=\pi/\sqrt{2C}$. Labels \([[N,K,d]]\) specify the numbers of data and logical qubits and the distance of the corresponding codes, respectively. \textbf{b,} Comparison between atom rearrangement and the Cat Bus with flag decoding under the same D2 error model. Points are Monte Carlo estimates; error bars denote one standard error propagated from binomial sampling and are smaller than the markers where not visible. Dotted lines mark the circuit-level threshold estimates, while the shaded regions indicate the uncertainty set by the sampled physical error rates. 
}
    \label{fig:threshold}
\end{figure}

To assess circuit-level performance, we perform two comparisons (Fig.~\ref{fig:threshold}). The first holds the Cat Bus circuit fixed while comparing the hardware-derived (HD) and two-qubit depolarizing (D2) noise models, thereby isolating the effect of noise structure. The second holds the D2 model fixed while comparing the Cat Bus and atom-rearrangement circuits, thereby isolating the architectural contribution. Both comparisons use a \(\mathrm{CZ}\)-limited error model in which single-qubit gates, SPAM and idling are ideal (Methods).

For the circuit-level comparisons, we fix \(\gamma/\kappa_1=18\) and \(\alpha=3=\alpha_{\mathrm{opt}}^{(1)}\), which minimizes the \(\mathrm{CZ}^{1}\) infidelity, and define the physical error rate as \(p\equiv\epsilon_{\mathrm{CZ}^{1}}\).
This convention matches \(p\) to the process infidelity assigned to each pairwise \(\mathrm{CZ}\) gate in the D2 model.
We use the same cat amplitude for the \(\mathrm{CZ}^{4}\) operation.
Accordingly, the HD model retains the native block structure of the channel derived in Sec.~\ref{sec: III CZ channel}, with leading error weights \(p\) and \(5p/2\) for the \(\mathrm{CZ}^{1}\) and \(\mathrm{CZ}^{4}\) blocks, respectively; whereas the D2 model applies the two-qubit depolarizing channel \(\mathcal D_2(p)\) independently after every pairwise \(\mathrm{CZ}\) gate (Methods). Logical failure rates are obtained using BP+OSD with sliding-window space--time decoding~\cite{Xu2024ConstantOverhead,Huang2024Increasing,Kang2025quits}. The detailed decoding settings are given in Methods.

Figure~\ref{fig:threshold}\textbf{a} gives a circuit-level threshold of \(p_{\mathrm{th,HD}}\approx0.80\%\) for the HD model, compared with \(p_{\mathrm{th,D2}}\approx0.46\%\) for D2. This increase reflects the biased, block-correlated structure of the HD channel relative to unstructured two-qubit depolarizing noise. Using \(p=\epsilon_{1,\min}=\pi/\sqrt{2C}\), the HD threshold corresponds to \(C_{\mathrm{th}}\approx7.8\times10^{4}\), within the regime targeted by recent advances in Rydberg-coupled microwave-cavity systems~\cite{Wilde2025Hybrid,Monica2025Cavity}.

Under the D2 model, Fig.~\ref{fig:threshold}\textbf{b} gives \(p_{\mathrm{th,rearrange}}\simeq0.55\%\) for atom rearrangement~\cite{Xu2024ConstantOverhead} and an approximate threshold \(p_{\mathrm{th,CB+flag}}\simeq0.72\%\) for the Cat Bus with \(X\)-flag decoding. The increase captures the circuit-level advantage provided by the Cat Bus architecture, including the flag information available from cat-ancilla measurements.

 In the sub-threshold regime, the HD model also exhibits systematic suppression with increasing code size. The LFR is well described by the empirical scaling law:
\begin{equation}
    \mathrm{LFR}_{\mathrm{Cat Bus, \, HD}}
    = 0.124 \left(
        \frac{p}{0.0079}
    \right)^{0.56N^{0.29}},
    \label{eq:subthreshold_scaling}
\end{equation}
where $N$ is the number of data qubits. Table~\ref{tab:subthreshold_scaling} summarizes the corresponding LFR estimates, including extrapolations to larger code sizes. At \(p=10^{-3}\), the predicted LFRs are \(7\times10^{-9}\) per cycle for \(N=10^{4}\) and \(7\times10^{-16}\) for \(N\simeq10^{5}\), corresponding to logical memory performance relevant to GigaQuOp-scale computation and cryptographic applications, respectively~\cite{Review2026}.

\begin{table}[t]
    \centering
    \caption{
    Logical failure rates per syndrome-extraction cycle predicted by
    the sub-threshold scaling law in Eq.~\eqref{eq:subthreshold_scaling} for representative members of
    the HGP-code family. Here, $N$ and $K$ denote the numbers of data
    and logical qubits, respectively, with a fixed encoding rate
    $K/N=1/25$.
    }
    \label{tab:subthreshold_scaling}

    \small
    \renewcommand{\arraystretch}{1.18}
    \setlength{\tabcolsep}{7pt}

    \begin{tabular}{@{}ccccc@{}}
        \toprule
        & \multicolumn{4}{c}{Code parameters $(N,K)$} \\
        \cmidrule(lr){2-5}
        $p$
        & $(225,9)$
        & $(1225,49)$
        & $(10000,400)$
        & $(102400,4096)$ \\
        \midrule
        $10^{-3}$
        & $5\times10^{-4}$
        & $1\times10^{-5}$
        & $7\times10^{-9}$
        & $7\times10^{-16}$ \\

        $10^{-4}$
        & $1\times10^{-6}$
        & $5\times10^{-10}$
        & $5\times10^{-17}$
        & $9\times10^{-32}$ \\
        \bottomrule
    \end{tabular}
\end{table}

\section{Discussion and Outlook}

We have introduced a hardware-efficient, shuttling-free architecture that replaces mechanical transport with a Cat Bus. We further show that this architecture enables faster syndrome-extraction cycle-time and a higher fault-tolerance threshold, yielding system-level
advantages for scalable fault-tolerant quantum architectures.

The Cat Bus architecture developed here is not restricted to neutral-atom systems and could also be adapted to superconducting platforms. Each atom could be replaced by a multilevel transmon whose auxiliary transition is resonantly coupled to the cat mode. Such a construction could realize an analogous interaction mechanism, enabling a native long-range $\mathrm{CZ}^{n}$ operation between the cat qubit and multiple transmons simultaneously. 
A recent demonstration of cat--transmon coupling and transmon-assisted bosonic syndrome extraction provides an experimental basis for exploring this direction~\cite{Putterman2025BosonicConcatenation}. 
Realizing the required selective transmon--cat coupling in a scalable device, together with characterizing the resulting transmon-specific error channels, remains an important direction for future work.

Moreover, the native parallel $\mathrm{CZ}^{n}$ operation provides a hardware-level quantum fan-out up to local Hadamard rotations. It can reduce the depth of selected multi-qubit operations from logarithmic to constant for circuits that admit fan-out constructions~\cite{Hoyer2005, Song2025}, within the fan-out supported by an individual Cat Bus.

Finally, extending the present quantum-memory benchmark to universal fault-tolerant computation will require logical gate constructions tailored to the Cat Bus architecture. A further direction is to interface the microwave cat modes with modular or microwave-to-optical interconnects~\cite{Niu2023,Meesala2024,Zhao2025}, which could extend the architecture beyond a single processor module, thus unlocking scalable pathways for modular quantum computing. More broadly, the co-design of non-local hardware connectivity, native gate parallelism and quantum error correction scheme offers a route towards resource-efficient fault-tolerant quantum computation.

\clearpage
\newpage

\bibliography{references}

\clearpage
\newpage

\section*{Methods}

\subsection*{Projected cat--atom interaction and finite-overlap corrections}

Each neutral atom contains computational states $\{\ket{0},\ket{1}\}$ and auxiliary Rydberg states $\{\ket{r},\ket{r'}\}$.
Throughout the Methods, $\sigma_x$, $\sigma_y$ and $\sigma_z$ denote Pauli operators in the Rydberg basis $\{\ket{r},\ket{r'}\}$, whereas $X$, $Y$ and $Z$ denote Pauli operators in the computational basis $\{\ket{0},\ket{1}\}$.
The operators $\Xcat$, $\Ycat$ and $\Zcat$ act on the cat-qubit manifold.
We choose the phase of the microwave mode such that $\alpha$ is real and positive.

Two-photon dissipation confines the cavity to $\mathcal C=\operatorname{span}\{\ket{\alpha},\ket{-\alpha}\}$~\cite{Guillaud2019RepetitionCat,Christopher2022FTQC,Putterman2025BosonicConcatenation}.
An orthonormal parity basis is
\begin{equation}
    \ket{\catp}=\mathcal N_+(\ket{\alpha}+\ket{-\alpha}),\qquad
    \ket{\catm}=\mathcal N_-(\ket{\alpha}-\ket{-\alpha}),
    \label{eq:methods_cat_basis}
\end{equation}
where $\mathcal N_\pm=[2(1\pm u)]^{-1/2}$ and $u=\braket{-\alpha}{\alpha}=e^{-2\alpha^2}$.
With $\Pcat=\ket{\catp}\bra{\catp}+\ket{\catm}\bra{\catm}$, projection of the cavity annihilation operator gives
\begin{align}
    \acat\equiv\Pcat\hat a\Pcat
    &=\frac{\alpha}{\sqrt{1-u^2}}
    \left(\Zcat-iu\Ycat\right)\nonumber\\
    &=\alpha(\Zcat-iu\Ycat)+\mathcal O(\alpha u^2).
    \label{eq:methods_projected_a}
\end{align}

For $n$ atoms simultaneously and resonantly coupled to the same microwave mode, the interaction is the sum of Jaynes--Cummings couplings
\begin{equation}
    \hat H_{\mathrm{JC}}=g\sum_{j=1}^{n}
    \left(\sigma_{+,j}\hat a+\sigma_{-,j}\hat a^\dagger\right),
    \label{eq:methods_HJC}
\end{equation}
where \(\sigma_{+,j}=\ket{r'}_j\bra r\), \(\sigma_{-,j}=\ket r_j\bra{r'}\), \(\sigma_{x,j}=\sigma_{+,j}+\sigma_{-,j}\) and \(\sigma_{y,j}=-i\ket r_j\bra{r'}+i\ket{r'}_j\bra r\).
Projection onto the stabilized cat manifold yields
\begin{align}
    \hat H_{\mathrm{CB}}
    &=\Pcat\hat H_{\mathrm{JC}}\Pcat\nonumber\\
    &=\frac{g\alpha}{\sqrt{1-u^2}}
    \sum_{j=1}^{n}
    \left(\sigma_{x,j}\Zcat-u\sigma_{y,j}\Ycat\right).
    \label{eq:methods_HCB}
\end{align}
At zeroth order in $u$, Eq.~\eqref{eq:methods_HCB} reduces to the commuting Hamiltonian used in Eq.~\eqref{eq:factorized_interaction}.

At finite \(u\), the second term in Eq.~\eqref{eq:methods_HCB} produces finite-overlap-induced coherent leakage with amplitude \(\mathcal O(u)\).
For a single target, this term gives a finite-overlap correction to the ideal gate; with multiple active targets, its dependence on the other targets gives multi-target crosstalk.
The resulting average infidelity is
\begin{equation}
    1-F_{\mathrm{avg}}=\bar c_n u^2+\mathcal O(u^4),
    \label{eq:methods_finite_overlap_infidelity}
\end{equation}
where \(\bar c_n\) contains a single-target contribution and, for \(n\geq2\), an additional crosstalk contribution.
The crosstalk contribution vanishes for \(n=1\), whereas \(\bar c_1=\pi^2/8\).
For \(n=4\) and \(\alpha=3\), Eq.~\eqref{eq:methods_finite_overlap_infidelity} gives \(1-F_{\mathrm{avg}}\simeq1.06\times10^{-15}\).
The complete finite-\(u\) derivation and the exact coefficients \(\bar c_n\) are given in the Supplementary Information.

\subsection*{Dissipative error channel of the cat--atom \texorpdfstring{\(\mathrm{CZ}^{n}\)}{CZn} gate}
\label{method:error_channel}

We derive the leading dissipative channel at \(u=0\), treating the preparation and return pulses as ideal and retaining terms linear in \(\kappa_1T_{\mathrm{int}}\) and \(\gamma T_{\mathrm{int}}\).
Corrections from the finite dissipative gap and mixed finite-overlap--dissipative terms are not included at this order.
For an addressed set \(\mathcal A\), the projected density matrix during the common interaction interval obeys
\begin{align}
    \dot\rho={}&-i[\hat H_{\mathrm{CB}}^{(0)},\rho]
    +\mathscr{D}[\sqrt{\kappa_1}\alpha\Zcat]\rho\nonumber\\
    &+\sum_{j\in\mathcal A}\sum_{s\in\{r,r'\}}
    \mathscr{D}[\sqrt{\gamma_s}\ket{\ell_s}_j\!\bra{s}_j]\rho,
    \label{eq:methods_projected_master}
\end{align}
where $\mathscr D[L]\rho=L\rho L^\dagger-\{L^\dagger L,\rho\}/2$.
The orthogonal sink states $\ket{\ell_r}$ and $\ket{\ell_{r'}}$ coarse-grain the decay products and their environmental records.
Because the two Rydberg levels have similar principal quantum numbers, we take $\gamma_r\simeq\gamma_{r'}\equiv\gamma$.

The leading channel follows from an interaction-picture expansion of Eq.~\eqref{eq:methods_projected_master}.
Single-photon loss gives one cat phase flip shared by all targets, whereas Rydberg decay gives one target-local joint cat--atom channel. We denote the cat phase-flip channel by
\begin{equation}
    \mathcal Z_{\mathrm{cat}}(q)[\rho]
    =(1-q)\rho+qZ_{\mathrm{cat}}\rho Z_{\mathrm{cat}}.
    \label{eq:methods_cat_phase_flip_channel}
\end{equation}
Here $q$ is the phase-flip probability.
For atom $j$, let $Z_j=\ket0_j\bra0-\ket1_j\bra1$ and $\rho_{11}^{(j)}=\bra1_j\rho\ket1_j$, where $\rho_{11}^{(j)}$ acts on the cat mode and all remaining atoms.
After the return pulse, the latter channel is
\begin{align}
    \mathcal E_{\mathrm{ryd}}^{(j)}[\rho]
    ={}&(1-p_{\mathrm{leak}})\rho
    +\frac{p_{\mathrm{leak}}}{2}(Z_j\rho+\rho Z_j)\nonumber\\
    &+p_{\mathrm{leak}}\ket{\ell_r}_j\bra{\ell_r}_j
    \otimes\rho_{11}^{(j)}\nonumber\\
    &+p_{\mathrm{leak}}\ket{\ell_{r'}}_j\bra{\ell_{r'}}_j
    \otimes\Zcat\rho_{11}^{(j)}\Zcat.
    \label{eq:methods_error_channel}
\end{align}
Here
\begin{equation}
    p_{Z_{\mathrm{cat}}}=\frac{\pi\kappa_1\alpha}{2g},
    \qquad
    p_{\mathrm{leak}}=\frac{\pi\gamma}{4g\alpha}.
    \label{eq:methods_error_probabilities}
\end{equation}
The two decay branches together remove population from an initially occupied $\ket1$ state with probability $2p_{\mathrm{leak}}=\gamma T_{\mathrm{int}}$, while their leading contribution to the process infidelity is $p_{\mathrm{leak}}$.
The interaction-picture Dyson derivation of Eq.~\eqref{eq:methods_error_channel} is given in the Supplementary Information.

For an addressed set $\mathcal A$, the native cat--atom $\mathrm{CZ}^{n}$ gate contains one cat phase-flip channel shared by all targets and one target-local Rydberg-decay channel for each atom:
\begin{equation}
    \mathcal E_{\mathrm{diss}}^{(\mathcal A)}
    =\mathcal{Z}_{\mathrm{cat}}(p_{Z_{\mathrm{cat}}})
    \circ
    \prod_{j\in\mathcal A}
    \mathcal E_{\mathrm{ryd}}^{(j)}(p_{\mathrm{leak}}).
    \label{eq:methods_native_block_channel}
\end{equation}

\newpage

\subsection*{Circuit-level noise models and decoding}

We use a $\mathrm{CZ}$-limited circuit-level model to isolate the propagation of entangling-gate errors through the syndrome-extraction circuits.
Single-qubit rotations, state preparation and measurement, and idling are ideal in this controlled benchmark.
For the hardware-derived (HD) model, we coarse-grain the channel in Eq.~\eqref{eq:methods_error_channel} into a cat phase-flip channel and a target-atom depolarizing channel.
Atomic leakage is represented by unheralded single-qubit depolarizing noise; retaining leakage or erasure information would define a different decoder model~\cite{Baranes2026Leveraging}.

At \(\gamma/\kappa_1=18\), the fan-out-specific optimum for \(\mathrm{CZ}^{4}\) would be \(\alpha_{\mathrm{opt}}^{(4)}=6\); the controlled benchmark instead uses the common operating point specified in the main text.
We define the elementary stochastic channels by
\begin{equation}
 \begin{aligned}
  \mathcal{Z}_{\mathrm{cat}}(q)[\rho]
  &=(1-q)\rho+qZ_{\mathrm{cat}}\rho Z_{\mathrm{cat}},\\
  \mathcal D_1^{(j)}(q)[\rho]
  &=(1-q)\rho
  +\frac{q}{3}
  \sum_{P_j\in\{X_j,Y_j,Z_j\}}P_j\rho P_j.
 \end{aligned}
 \label{eq:methods_elementary_noise_channels}
\end{equation}
For a native block $b=(c_b,\mathcal A_b)$, with Cat Bus $c_b$ and simultaneously addressed target set $\mathcal A_b$, the HD channel is
\begin{equation}
 \mathcal E_{\mathrm{HD}}^{(b)}(p)
 =
 \mathcal{Z}_{\mathrm{cat}}^{(c_b)}(p/2)
 \circ
 \prod_{j\in\mathcal A_b}\mathcal D_1^{(j)}(p/2).
 \label{eq:methods_HD_block_channel}
\end{equation}
The cat channel acts once during the common interaction interval, whereas the atomic channel acts independently on each target.
The simulations contain native $\mathrm{CZ}^{1}$ and $\mathrm{CZ}^{4}$ blocks.
Their leading error weights are \(p\) and \(5p/2\), respectively.

The ideal multi-target operation may be serialized as a product of pairwise $\mathrm{CZ}$ gates in the stabilizer circuit, but its HD noise retains the native block grouping shown in Table~\ref{tab:methods_noise_models}.
By contrast, the comparison model applies $\mathcal D_2(p)$ independently after every $\mathrm{CZ}$ gate.

The noisy stabilizer circuits and detector error models are generated with Stim~\cite{Gidney2021Stim}.
Before the first noisy cycle, all data qubits receive a phenomenological depolarizing channel of probability \(p_{\mathrm r}=5p\), representing the residual data error passed from the preceding single-shot decoding block~\cite{Xu2024ConstantOverhead}.
We decode non-overlapping $(3,3)$ space--time windows~\cite{Xu2024ConstantOverhead,Huang2024Increasing,Kang2025quits}.
Each three-cycle block is decoded by min-sum belief propagation (BP), and its inferred update is hard-committed.
The final noiseless data readout supplies the temporal boundary, after which BP with ordered-statistics decoding (BP+OSD) is applied using OSD-e at order 15~\cite{Grospellier2021combininghardsoft,Panteleev2021GoodLDPC}.
The maximum BP iteration counts are
\begin{equation}
 I_{\max}^{(0)}=\frac{N}{5},
 \qquad
 I_{\max}^{(X)}=\frac{2.75N}{5},
 \label{eq:methods_BP_iterations}
\end{equation}
for syndrome-only and syndrome-plus-$X$-flag decoding, respectively.
The min-sum scaling factor is 0.625 for the HD-versus-D2 comparison and 0.9 for the Cat Bus-versus-rearrangement comparison.

Cat flags are omitted in the HD-versus-D2 comparison so that the detector set is identical for the two noise models.
For the Cat Bus-versus-rearrangement comparison under D2 noise, the Cat Bus decoder additionally receives the $X$-basis cat flags; the $Z$-basis cat flags are not included.
We simulate \(4.5d\) syndrome-extraction cycles for the HD-versus-D2 comparison and \(3d\) cycles for the architecture comparison, as determined from the finite-cycle relaxation analysis in the Supplementary Information.
For a memory experiment containing \(m\) syndrome-extraction cycles, let \(P_L\) denote the probability that at least one logical observable is flipped after final decoding. We report the block logical failure rate per cycle as
\begin{equation}
 \mathrm{LFR}=1-(1-P_L)^{1/m}.
 \label{eq:methods_LFR}
\end{equation}
The circuit-level threshold is the critical value of the corresponding model parameter \(p\) below which the logical failure rate per cycle decreases with increasing code distance.
Sampling limits, finite-cycle fits, flag-information analysis and subthreshold fitting are detailed in the Supplementary Information.

\begin{table}[t]
 \caption{\label{tab:methods_noise_models}
 Circuit-level entangling-gate noise models.
 Each listed channel is applied after the corresponding gate or block.
 For the HD model, \(p=\epsilon_{\mathrm{CZ}^{1}}\) is the process infidelity of the \(\mathrm{CZ}^{1}\) gate; for the D2 model, \(p\) is the process infidelity of each noisy \(\mathrm{CZ}\) gate.
 Here \(\mathcal P_2\) denotes the two-qubit Pauli group, and \(\mathcal P_2^\ast\equiv\mathcal P_2\setminus\{I\}\) denotes its non-identity elements.
 HD error weights are given to first order in \(p\).}
 \centering
 \small
 \setlength{\tabcolsep}{2pt}
 \begin{tabular*}{\columnwidth}{@{\extracolsep{\fill}}lll@{}}
  \toprule
  \parbox[t]{0.11\columnwidth}{Model}
  & \parbox[t]{0.17\columnwidth}{Operation}
  & \parbox[t]{0.66\columnwidth}{Error channel} \\
  \midrule
  \multirow{2}{0.11\columnwidth}{\centering\raisebox{-7pt}[0pt][0pt]{HD}}
  & \parbox[t]{0.17\columnwidth}{$\mathrm{CZ}^{1}$}
  & \parbox[t]{0.66\columnwidth}{\(\displaystyle
    \mathcal{Z}_{\mathrm{cat}}(p/2)\circ\mathcal D_1^{(j)}(p/2)
   \)} \\[5pt]
  & \parbox[t]{0.17\columnwidth}{$\mathrm{CZ}^{4}$}
  & \parbox[t]{0.66\columnwidth}{\(\displaystyle
    \mathcal{Z}_{\mathrm{cat}}(p/2)\circ\prod_{j=1}^{4}\mathcal D_1^{(j)}(p/2)
   \)} \\[5pt]
  \midrule
  \parbox[t]{0.11\columnwidth}{D2}
  & \parbox[t]{0.17\columnwidth}{$\mathrm{CZ}$}
  & \parbox[t]{0.66\columnwidth}{\(\displaystyle
    \mathcal D_2(p)[\rho]
    =(1-p)\rho+\frac{p}{15}
    \sum_{P\in\mathcal P_2^\ast}P\rho P
   \)} \\
  \bottomrule
 \end{tabular*}
\end{table}

\newpage

\subsection*{Optimality of the two-dimensional star-based scheduling}

Consider an HGP code constructed from two classical Tanner graphs, $G_{\mathrm H}$ and $G_{\mathrm V}$~\cite{Tillich2009HGP}. For an individual check measurement of fixed type $\sigma\in\{X,Z\}$, the product structure partitions the required ancilla--data interactions into horizontal and vertical sectors. These sectors define independent scheduling problems, but the Cat-Bus architecture requires their interactions to be executed in separate intervals. Their depths therefore add:
\begin{equation}
    T_{\sigma}
    =T_{\mathrm H,\sigma}+T_{\mathrm V,\sigma}.
    \label{eq:orientation_depth_sum}
\end{equation}

Within orientation $f\in\{\mathrm H,\mathrm V\}$, the product construction generates topologically identical copies of the corresponding classical Tanner graph $G_f$. Because the copies are assigned to distinct Cat Buses, they can be executed concurrently.
The depth of each orientation is therefore determined by the one-dimensional star-based scheduling of $G_f$.

Let $\tau(G_f)$ denote the minimum vertex-cover number of $G_f$. In any $T$-step star-based scheduling of one copy, the selected star centers must collectively cover every edge and therefore form a vertex cover. Thus, $T\geq\tau(G_f)$. The MVC construction in Algorithm~\ref{alg:star_decomposition} attains this lower bound and can be applied concurrently to every copy. Hence,
\begin{equation}
    T_{f,\sigma}^{\min}=\tau(G_f),
    \qquad
    f\in\{\mathrm H,\mathrm V\},
    \quad \sigma\in\{X,Z\}.
    \label{eq:orientation_optimal_depth}
\end{equation}
Combining the two orientations gives
\begin{equation}
    T_{\sigma}^{\min}
    =\tau(G_{\mathrm H})+\tau(G_{\mathrm V}),
    \qquad \sigma\in\{X,Z\}.
    \label{eq:general_optimal_2d_star_scheduling}
\end{equation}

In this work, we consider HGP codes constructed from two identical classical Tanner graphs, such that $G_{\mathrm H}=G_{\mathrm V}=G$. For a \((\Delta_{\mathrm{C}},\Delta_{\mathrm{V}})\)-biregular Tanner graph, a maximum matching saturates all \(n_{\mathrm C}\) check nodes  ~\cite{LovaszPlummer1986}, giving \(\tau(G)=n_{\mathrm C}\).
Equation~\eqref{eq:general_optimal_2d_star_scheduling} therefore reduces to
\begin{equation}
    T_{\sigma}^{\min}=2\tau(G)=2n_{\mathrm C}.
    \label{eq:identical_code_optimal_depth}
\end{equation}

This result concerns an individual $X$- or $Z$-check measurement. For repeated syndrome extraction, the star-based construction can be combined with the pipelined ordering of Xu \emph{et al.}~\cite{Xu2024ConstantOverhead}. Compatible scheduling blocks from successive measurement rounds can be interleaved schematically as
\begin{equation}
    X_{\mathrm H}^{(1)}
    \longrightarrow
    \bigl[
        X_{\mathrm V}^{(1)}
        \parallel
        Z_{\mathrm V}^{(1)}
    \bigr]
    \longrightarrow
    \bigl[
        Z_{\mathrm H}^{(1)}
        \parallel
        X_{\mathrm H}^{(2)}
    \bigr]
    \longrightarrow\cdots ,
    \label{eq:pipelined_star_scheduling}
\end{equation}
where the superscript labels the measurement round and $\parallel$ denotes compatible blocks executed concurrently. For $r$ repeated rounds, this pipelined ordering reduces the scheduling depth from $4r n_{\mathrm C}$ to $(2r+2)n_{\mathrm C}$. The additional $2n_{\mathrm C}$ corresponds to two boundary blocks, each of depth $n_{\mathrm C}$, that cannot be overlapped with operations from an adjacent measurement round.

\newpage
\subsection*{Cat-Mediated Syndrome Extraction Circuit}

We construct the star scheduling from a MVC of the underlying classical Tanner graph. Using a maximum matching together with K\"{o}nig's theorem~\cite{LovaszPlummer1986}, we identify a minimum set of star centers whose incident edges cover the entire graph; each resulting star defines one step of the one-dimensional classical scheduling. The detailed procedure is given in Algorithm~\ref{alg:star_decomposition}.

For the $(3,4)$-biregular Tanner graphs considered here, the maximum matching saturates all check nodes, and the resulting MVC is the full set of check nodes. Each check node therefore serves as the center of a degree-4 star.

\begin{algorithm}[H]
    \caption{1D classical scheduling scheme for arbitrary bipartite graph}
    \label{alg:star_decomposition}
    \KwIn{$H \in \{0,1\}^{m \times n}$: classical parity-check matrix.}
    \KwOut{$S = (S_1, S_2, \dots, S_T)$: an ordered list of stars. Each $S_t$ is a set of check-bit edges that can be scheduled together.}
    \BlankLine
    
    \SetKwProg{Fn}{Function}{:}{}
    \SetKwFunction{FStarDecomp}{StarDecomposition}
    
    \Fn{\FStarDecomp{$H$}}{
        Construct a bipartite graph $G = (C \cup V, E)$, where $C = \{c_1, \dots, c_m\}$ are check nodes, $V = \{v_1, \dots, v_n\}$ are bit nodes, and $(c_i, v_j) \in E$ iff $H[i,j] = 1$.\;
        Find a maximum matching $M$ of $G$.\;
        Let $U$ be the set of unmatched check nodes in $C$.\;
        Starting from $U$, find the set $Z$ of vertices reachable by alternating paths: from $C$ to $V$ along unmatched edges, and from $V$ to $C$ along matched edges.\;
        By K\"{o}nig's theorem, compute a minimum vertex cover $K = (C \setminus Z) \cup (V \cap Z)$.\;
        Initialize $S \leftarrow \emptyset$ and $E_{\text{used}} \leftarrow \emptyset$.\;
        
        \For{each center node $u \in K$}{
            $S_u \leftarrow \emptyset$\;
            \For{each neighbor $w$ of $u$}{
                \If{edge $(u,w) \notin E_{\text{used}}$}{
                    Add $(u,w)$ to $S_u$\;
                    Add $(u,w)$ to $E_{\text{used}}$\;
                }
            }
            \If{$S_u \neq \emptyset$}{
                Append $S_u$ to $S$\;
            }
        }
        \KwRet $S$\;
    }
\end{algorithm}
Using the tensor-product structure of HGP codes, the one-dimensional star scheduling is lifted onto the two-dimensional qubit array, generating horizontal and vertical scheduling steps for the check ancilla--data interactions. Algorithm~\ref{alg:product_circuit} specifies this product-lifting procedure for both $X$- and $Z$-check rounds.
\\
\begin{algorithm}[H]
    \label{alg:product_circuit}
    \caption{Parallel 2D product scheduling for HGP codes}
    \KwIn{
        $H_X$ or $H_Z$: quantum check matrix of the HGP code.\newline
        $\tau \in \{X, Z\}$: check type to be measured.\newline
        $n_C, n_V$: numbers of check and bit nodes in the underlying classical code.\newline
        $\text{pos}_Q$: map from 2D lattice positions to data-qubit indices.\newline
        $\text{pos}_A^\tau$: map from 2D lattice positions to $\tau$-check ancilla indices.
    }
    \KwOut{$P = (P_1, P_2, \dots, P_L)$: ordered product-circuit scheduling. Each $P_l$ is a set of ancilla-data interactions executable in parallel.}
    \BlankLine
    
    \SetKwProg{Fn}{Function}{:}{}
    \SetKwFunction{FProductscheduling}{Productscheduling}
    
    \Fn{\FProductscheduling{$H_\tau$, $\tau$}}{
        Extract the underlying classical parity-check matrix $H_{\text{class}}$ from $H_\tau$\;
        
        $S \leftarrow \text{StarDecomposition}(H_{\text{class}})$\;
        $P \leftarrow \emptyset$\;
        
        \For{each star $S_t$ in $S$}{
            $H_t^X, V_t^X, H_t^Z, V_t^Z\leftarrow \emptyset$ \tcp*{horizontal/vertical $X$/$Z$ product interactions}
            
            \If{$\tau = X$}{
                \For{each classical edge $(c, v) \in S_t$}{
                    \For{each product coordinate $y = n_V, \dots, n_V + n_C - 1$}{
                        Add interaction (data qubit at position $(y, c)$, $X$-ancilla at position $(y, v + n_C)$) to $H_t^X$\;
                    }
                    \For{each product coordinate $x = n_C, \dots, n_C + n_V - 1$}{
                        Add interaction ($X$-ancilla at position $(c + n_V, x)$, data qubit at position $(v, x)$) to $V_t^X$\;
                    }
                }
                }

            \If{$\tau = Z$}{
            \For{each classical edge $(c, v) \in S_t$}{
                \For{each product coordinate $y = 0, \dots, n_V - 1$}{
                    Add interaction ($Z$-ancilla at position $(y, c)$, data qubit at position $(y, v + n_C)$) to $H_t^Z$\;
                }
                \For{each product coordinate $x = 0, \dots, n_C - 1$}{
                    Add interaction (data qubit at position $(c + n_V, x)$, $Z$-ancilla at position $(v, x)$) to $V_t^Z$\;
                }
            }
            }
            Append $H_t^X, V_t^X, H_t^Z, V_t^Z$ to $P$\;
        }
        \KwRet $P$\;
    }
\end{algorithm}
Each ancilla--data pair is ordered with the bus-selecting atom first: $(X\text{-ancilla},\text{data})$ for $V_t^X$, $(\text{data},X\text{-ancilla})$ for $H_t^X$, $(Z\text{-ancilla},\text{data})$ for $H_t^Z$, and $(\text{data},Z\text{-ancilla})$ for $V_t^Z$. In all cases, the first entry determines the aligned row or column cat ancilla used to mediate the interaction, whereas the second entry denotes an atomic target coupled through that Cat Bus. Each scheduling step is subsequently converted into an explicit cat-mediated circuit by grouping all pairs that share the same first entry. Algorithm~\ref{alg:cat_xvertical} illustrates this circuit transformation for an $X$-vertical scheduling step.
\begin{algorithm}[H]
    \caption{Transfer an atom--atom $\mathrm{CZ}^{n}$ gate to the equivalent cat-mediated circuit}
    \label{alg:cat_xvertical}
    \KwIn{
        $T_t$: the $t$-th time step of $V_t^X$. It is a set of ordered pairs $(a, q)$ that are scheduled to be implemented in the same time step, where $a = X$-check ancilla, $q = \text{data qubit}$.\newline
        $C$: column cat ancilla qubits.
    }
    \KwOut{A parallel cat-mediated implementation of all interactions in $T_t$.}
    \BlankLine
    
    \SetKwProg{Sub}{Subroutine}{:}{}
    \SetKwFunction{FApplyCATLayer}{ApplyCat\_XVertical}
    
    \Sub{\FApplyCATLayer{$T_t$}}{
        Let $A_t$ be the set of distinct $X$-check ancillas appearing in $T_t$\;
        
        \For{each $a \in A_t$}{
            Choose the column cat ancilla $c(a)$ aligned with $a$\;
            Prepare $c(a)$ in state $|+\rangle$\;
        }
        
        \BlankLine
        \textbf{In parallel:}\;
        \Indp
            Apply $\text{CZ}$ between each $c(a)$ and its $X$-check ancilla $a$\;
        \Indm
        
        \BlankLine
        Apply $H$ to all selected cat ancillas\;
        
        \BlankLine
        \textbf{In one parallel $\mathrm{CZ}^{n}$ layer:}\;
        \For{each $a \in A_t$}{
            Apply $\mathrm{CZ}^{n}$ from $c(a)$ to all data qubits $q$ such that $(a, q) \in T_t$\;
        }
        
        \BlankLine
        Apply $H$ to all selected cat ancillas\;
        
        \BlankLine
        \textbf{In parallel:}\;
        \Indp
            Apply $\text{CZ}$ between each $c(a)$ and its $X$-check ancilla $a$\;
        \Indm
        
        \BlankLine
        Measure all selected cat ancillas in the $X$ basis\;
        Record the cat measurement outcomes as flag outcomes\;
    }
\end{algorithm}

Algorithm~\ref{alg:cat_xvertical} gives the cat-mediated circuit implementation of a single $X$-vertical scheduling step. The corresponding routines for the $X$-horizontal, $Z$-horizontal, and $Z$-vertical steps are defined analogously by exchanging the check type and bus orientation while preserving the same circuit structure.

To implement a complete syndrome-extraction cycle, we first prepare the $X$-check ancillas in $|+\rangle$, apply Hadamard gates to the data qubits, and execute the $X$-horizontal and $X$-vertical scheduling steps using the corresponding \textsc{ApplyCat} routines.
We then undo the data-qubit basis rotation, prepare the $Z$-check ancillas in $|+\rangle$, and execute the corresponding $Z$-horizontal and $Z$-vertical steps.
Finally, all check ancillas are measured in the $X$ basis, yielding the syndrome outcomes $m_X$ and $m_Z$.

This procedure completes the compilation of the one-dimensional star scheduling into the hardware-level cat-mediated syndrome-extraction circuit. For the $(3,4)$-biregular Tanner graphs considered here, each check round contains $n_{\mathrm C}$ horizontal and $n_{\mathrm C}$ vertical scheduling steps. A complete syndrome-extraction cycle, consisting of sequential $X$- and $Z$-check rounds, therefore requires $4n_{\mathrm C}$ scheduling steps.\\

\noindent \textbf{Data availability} \\
The data that support the findings of this article are openly available. \\

\noindent \textbf{Code availability}\\
All codes used to generate the numerical results and
figures are available upon request. \\

\noindent \textbf{Acknowledgments} \\
We thank Jianwei Pan, Hongzheng Zhao and Lei Feng for helpful discussions. This work is supported by National Program on Key Basic Research Project of China (Grant No. 2021YFA1400900), the Innovation Program for Quantum Science and Technology of China (Grant No. 2024ZD0300100), Shanghai Municipal Science and Technology Major Project (Grant No. 26LZ0500100, 25TQ003, 24LZ1400900, 24LZ1401600, 24DP2600100). \\

\noindent \textbf{Author contributions} \\
X.L. conceived and supervised the project. Y.C. and X.Y. jointly developed the Cat Bus architecture and syndrome-extraction scheme, performed the theoretical analysis and numerical simulations, and wrote the manuscript. Y.M. contributed to the design of the cat--atom gate protocol. Z.Z. and S.C. contributed to the implementation of the quantum-error-correction simulation code. All authors discussed the results and reviewed the manuscript.\\

\noindent \textbf{Competing interests} \\
There are no competing interests. \\

\end{document}


\title{Supplementary Information for\\
Fault-tolerant quantum computing with a microwave Cat Bus}

\author{Yanyan Chen}\email{yanyanchen235@gmail.com}
\thanks{These authors contributed equally to this work.}

\author{Xinyang Yu}\email{yuxy18@fudan.edu.cn}
\thanks{These authors contributed equally to this work.}

\author{Yueyang Min}
\author{Zhihao Zhang}
\author{Shuaifan Cao}

\author{Xiaopeng Li}\email{xiaopeng\_li@fudan.edu.cn}

\date{\today}

\begin{abstract}
This Supplementary Information provides the analytical derivations and numerical methods supporting the main text.
Section~\ref{Sec:gate_protocol} develops the cat-code representation and native cat--atom $\mathrm{CZ}^{n}$ protocol, including constant-depth entangling gates between arbitrary pairs of atoms.
Sections~\ref{Sec:dissipative_error} and~\ref{Sec:crosstalk} derive the dissipative error channel and finite-overlap-induced multi-target crosstalk, respectively.
Section~\ref{Sec:syndrome_extraction} presents the minimum-vertex-cover scheduling, circuit implementation and cycle-time comparison for Cat-Bus-mediated syndrome extraction.
Section~\ref{Sec:simulations_decoding} specifies the circuit-level noise models, finite-cycle protocol, use of cat-flag information and threshold and subthreshold analyses.
\end{abstract}

\maketitle

\tableofcontents

\section{Cat-code representation and native gate protocol}
\label{Sec:gate_protocol}

\subsection{Stabilized cat manifold and projected operators}

We choose the phase of the microwave mode such that $\alpha$ is real and positive.
Each cavity mode is stabilized by the two-photon jump operator
\begin{equation}
    \hat{L}_2=\sqrt{\kappa_2}\left(\hat{a}^2-\alpha^2\right),
\end{equation}
whose steady-state manifold is $\mathcal{C}=\operatorname{span}\{\ket{\alpha},\ket{-\alpha}\}$~\cite{Guillaud2019RepetitionCat,Christopher2022FTQC,Putterman2025BosonicConcatenation}.
The overlap of the coherent components is
\begin{equation}
    u\equiv\braket{-\alpha}{\alpha}=e^{-2\alpha^2}.
\end{equation}
An orthonormal basis for $\mathcal{C}$ is formed by the even- and odd-parity cat states
\begin{align}
    \ket{\catp}
    &=\mathcal{N}_+\left(\ket{\alpha}+\ket{-\alpha}\right),
    \label{Seq:even_cat}\\
    \ket{\catm}
    &=\mathcal{N}_-\left(\ket{\alpha}-\ket{-\alpha}\right),
    \label{Seq:odd_cat}
\end{align}
where $\mathcal{N}_\pm=[2(1\pm u)]^{-1/2}$.
The projector onto the stabilized manifold is
\begin{equation}
    \Pcat=\ket{\catp}\bra{\catp}+\ket{\catm}\bra{\catm}.
\end{equation}

Within this basis, we define the logical Pauli operators
\begin{align}
    \Xcat
    &=\ket{\catp}\bra{\catp}-\ket{\catm}\bra{\catm},\nonumber\\
    \Ycat
    &=i\ket{\catp}\bra{\catm}-i\ket{\catm}\bra{\catp},\nonumber\\
    \Zcat
    &=\ket{\catp}\bra{\catm}+\ket{\catm}\bra{\catp}.
\end{align}
They satisfy $\Xcat^2=\Ycat^2=\Zcat^2=\Pcat$ and $[\Xcat,\Ycat]=2i\Zcat$ within $\mathcal{C}$.

Projection of the bosonic annihilation operator gives $\acat\equiv\Pcat\hat{a}\Pcat$.
Its action on the parity basis is
\begin{align}
    \hat{a}\ket{\catp}
    &=\alpha\sqrt{\frac{1-u}{1+u}}\ket{\catm},&
    \hat{a}\ket{\catm}
    &=\alpha\sqrt{\frac{1+u}{1-u}}\ket{\catp}.
\end{align}
Consequently,
\begin{align}
    \acat
    &=\frac{\alpha}{\sqrt{1-u^2}}
    \left(\Zcat-iu\Ycat\right)
    \label{Seq:acat}\\
    &=\alpha\left(\Zcat-iu\Ycat\right)
    +\mathcal{O}(\alpha u^2).
\end{align}
The residual non-commutativity of the projected ladder operators is
\begin{equation}
    [\acat,\acat^\dagger]
    =\frac{4\alpha^2u}{1-u^2}\Xcat
    =4\alpha^2u\Xcat+\mathcal{O}(\alpha^2u^3).
\end{equation}
Thus, $\acat$ is Hermitian at zeroth order in $u$, whereas finite coherent-state overlap produces an exponentially suppressed non-Hermitian correction.

\subsection{Projected cat--atom Hamiltonian}

Each neutral atom contains computational states $\{\ket{0},\ket{1}\}$ and auxiliary Rydberg states $\{\ket{r},\ket{r'}\}$.
During the cat--atom interaction, the transition $\ket{r}\leftrightarrow\ket{r'}$ is resonant with the cavity mode, while the computational states are uncoupled.
For atom $j$, we define
\begin{align}
    \sigma_{+,j}
    &=\ket{r'}_j\bra{r},
    &
    \sigma_{-,j}
    &=\ket{r}_j\bra{r'},\nonumber\\
    \sigma_{x,j}
    &=\ket{r}_j\bra{r'}+\ket{r'}_j\bra{r},\nonumber\\
    \sigma_{y,j}
    &=-i\ket{r}_j\bra{r'}+i\ket{r'}_j\bra{r}.
\end{align}
Here, $\sigma_{x,j}$ and $\sigma_{y,j}$ are Pauli operators on the Rydberg doublet; $X_j$, $Y_j$ and $Z_j$ are reserved for the computational basis.

For $n$ atoms coupled to one cavity with single-atom vacuum coupling strength $g$, the sum of simultaneous resonant Jaynes--Cummings couplings is
\begin{equation}
    \hat{H}_{\mathrm{JC}}
    =g\sum_{j=1}^{n}
    \left(\sigma_{+,j}\hat{a}+\sigma_{-,j}\hat{a}^\dagger\right).
\end{equation}
In the dissipative-Zeno regime $\kappa_2\gg g$, projection onto the stabilized cat manifold yields
\begin{align}
    \hat{H}_{\mathrm{CB}}
    &=\Pcat\hat{H}_{\mathrm{JC}}\Pcat\nonumber\\
    &=\frac{g\alpha}{\sqrt{1-u^2}}
    \sum_{j=1}^{n}
    \left(\sigma_{x,j}\Zcat-u\sigma_{y,j}\Ycat\right)
    \label{Seq:Heff}\\
    &\equiv\frac{g\alpha}{\sqrt{1-u^2}}
    \left(\hat{H}_0+u\hat{V}\right),
    \nonumber
\end{align}
where
\begin{equation}
    \hat{H}_0=\sum_{j=1}^{n}\sigma_{x,j}\Zcat,
    \qquad
    \hat{V}=-\sum_{j=1}^{n}\sigma_{y,j}\Ycat.
\end{equation}
At $u=0$, all constituent interactions commute:
\begin{equation}
    [\sigma_{x,j}\Zcat,\sigma_{x,k}\Zcat]=0.
\end{equation}
The corresponding evolution therefore factorizes as
\begin{equation}
    \hat{U}_0(t)
    =\prod_{j=1}^{n}
    \exp\left(-ig\alpha t\,\sigma_{x,j}\Zcat\right).
\end{equation}
Finite overlap adds the noncommuting order-$u$ term $\hat{V}$.
Its effect on the complete gate and the associated multi-target crosstalk are derived in Section~\ref{Sec:crosstalk}.

\subsection{Native one-to-many \texorpdfstring{$\mathrm{CZ}^{n}$}{} protocol}

\begin{figure}[htbp]
    \centering
    \includegraphics[width=0.95\linewidth]{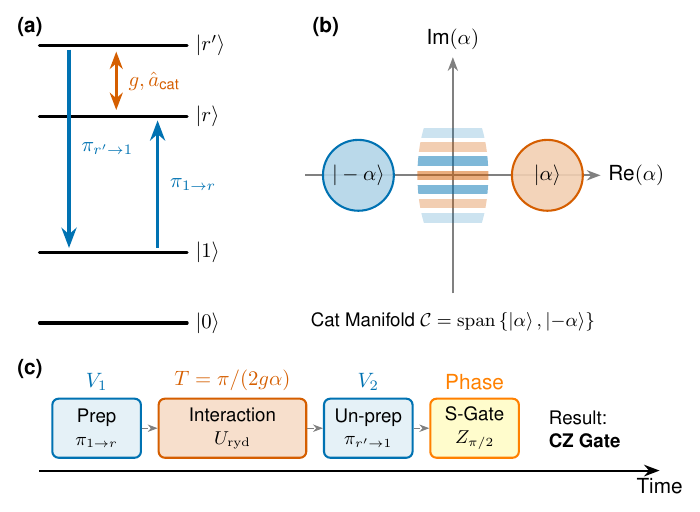}
    \caption{\textbf{Native cat--atom controlled-phase sequence.}
    A resonant $\pi$ pulse maps the addressed atomic component $\ket{1}$ to $\ket{r}$, while $\ket{0}$ remains dark.
    The cat--atom interaction is then applied for $T_{\mathrm{int}}=\pi/(2g\alpha)$, followed by a return pulse from $\ket{r'}$ to $\ket{1}$ and a virtual local phase correction $S$.
    In the ideal $u\rightarrow0$ limit, the sequence implements a controlled-$Z$ gate between the atom and the Cat Bus.}
    \label{fig:gate_sequence}
\end{figure}

For one addressed atom, the preparation and return pulses are represented by
\begin{align}
    \hat{V}_1
    &=\ket{r}\bra{1}+\ket{1}\bra{r}
    +\ket{0}\bra{0}+\ket{r'}\bra{r'},\nonumber\\
    \hat{V}_2
    &=\ket{r'}\bra{1}+\ket{1}\bra{r'}
    +\ket{0}\bra{0}+\ket{r}\bra{r}.
\end{align}
During the interaction interval, the relevant atomic space is $\operatorname{span}\{\ket{0},\ket{r},\ket{r'}\}$.
Defining $P_0=\ket{0}\bra{0}$ and $Q=\ket{r}\bra{r}+\ket{r'}\bra{r'}$, the leading-order propagator is
\begin{equation}
    e^{-i\theta\sigma_x\Zcat}
    =P_0+\cos\theta\,Q-i\sin\theta\,\sigma_x\Zcat.
    \label{Seq:single_atom_rotation}
\end{equation}
At
\begin{equation}
    T_{\mathrm{int}}=\frac{\pi}{2g\alpha}
\end{equation}
it reduces to
\begin{equation}
    \hat{U}_{\mathrm{int}}^{(0)}
    =P_0-i\sigma_x\Zcat.
    \label{Seq:single_atom_ideal_interaction}
\end{equation}
The first term in Eq.~\eqref{Seq:single_atom_ideal_interaction} leaves the dark state $\ket{0}$ invariant.
In the $u\rightarrow0$ limit, the second maps $\ket{r}\ket{\pm\alpha}\rightarrow\mp i\ket{r'}\ket{\pm\alpha}$.

After the return pulse, a virtual phase correction
\begin{equation}
    \hat{S}
    =\ket{0}\bra{0}+i\ket{1}\bra{1}
    +\ket{r}\bra{r}+\ket{r'}\bra{r'}
\end{equation}
removes the state-independent phase.
With $\hat{P}_{\mathcal{S}}=\ket{0}\bra{0}+\ket{1}\bra{1}$, the ideal logical action is
\begin{equation}
    \hat{U}_{\mathrm{CZ}}
    =\hat{P}_{\mathcal{S}}\hat{S}\hat{V}_2
    \hat{U}_{\mathrm{int}}^{(0)}\hat{V}_1\hat{P}_{\mathcal{S}}
    =\operatorname{diag}(1,1,1,-1)
\end{equation}
in the ordered basis $\{\ket{0}\ket{\alpha},\ket{0}\ket{-\alpha},\ket{1}\ket{\alpha},\ket{1}\ket{-\alpha}\}$ at $u\rightarrow0$.

The same pulse sequence can be applied simultaneously to an addressed set $\mathcal{A}_i$ coupled to cavity $i$.
Define
\begin{align}
    \hat{P}_{\mathcal{S}}^{(\mathcal{A}_i)}
    &=\prod_{k\in\mathcal{A}_i}\hat{P}_{\mathcal{S},k},
    \nonumber\\
    \hat{V}_{\nu}^{(\mathcal{A}_i)}
    &=\prod_{k\in\mathcal{A}_i}\hat{V}_{\nu,k},
    \qquad \nu\in\{1,2\},
    \nonumber\\
    \hat{S}^{(\mathcal{A}_i)}
    &=\prod_{k\in\mathcal{A}_i}\hat{S}_k.
\end{align}
The factorized interaction then gives
\begin{align}
    \hat{U}_{\mathrm{CZ}^{n}}^{(i)}
    &=
    \hat{P}_{\mathcal{S}}^{(\mathcal{A}_i)}
    \hat{S}^{(\mathcal{A}_i)}
    \hat{V}_{2}^{(\mathcal{A}_i)}
    \hat{U}_0(T_{\mathrm{int}})
    \hat{V}_{1}^{(\mathcal{A}_i)}
    \hat{P}_{\mathcal{S}}^{(\mathcal{A}_i)}
    \nonumber\\
    &=\prod_{k\in\mathcal{A}_i}
    \mathrm{CZ}_{\mathrm{cat}_i,k}.
\end{align}
All addressed targets share the same interaction interval, so the ideal gate duration is independent of $|\mathcal{A}_i|$.

\begin{figure}[htbp]
    \centering
    \makebox[\linewidth][c]{%
        \hspace*{-0.085\linewidth}%
        \includegraphics[width=0.8\linewidth]{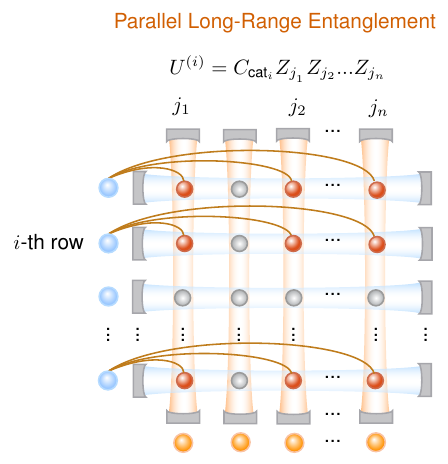}%
        \hspace*{0.085\linewidth}%
    }
    \caption{\textbf{One-to-many gates underlying all-to-all connectivity.}
    Each horizontal blue resonator and vertical orange resonator intersects every atom in its row or column.
    The highlighted row shows one Cat Bus implementing $\prod_{k\in\mathcal{A}_i}\mathrm{CZ}_{\mathrm{cat}_i,k}$ simultaneously on the addressed atoms (red).
    Unaddressed atoms remain idle (grey).
    Row and column buses provide the one-to-many primitives required for non-local connectivity across the static array.
    Distinct buses can operate concurrently when their addressed sets are compatible.}
    \label{fig:czn}
\end{figure}

\subsection{Constant-depth entangling gates between arbitrary pairs of atoms}

The crossed-bus geometry directly couples atoms that share a row or column.
All-to-all connectivity additionally requires a constant-depth protocol for atoms that share neither bus.

Consider two atoms \(A\) and \(B\) that do not share a Cat Bus.
The atom \(M\), located at the intersection of their row and column, couples to the corresponding buses \(C_{\mathrm r}\) and \(C_{\mathrm c}\).
Both buses are initialized in \(\ket0\), whereas the state of \(M\) is arbitrary.
Local Hadamard rotations convert the native cat--atom \(\mathrm{CZ}\) gate into a CNOT in either direction.
The gate \(\mathrm{CNOT}_{A\rightarrow B}\) can therefore be implemented in four entangling layers:
\begin{equation}
    \begin{array}{c|l}
        1 & \mathrm{CNOT}_{A\rightarrow C_{\mathrm r}}
            \ \parallel\
            \mathrm{CNOT}_{M\rightarrow C_{\mathrm c}}\\
        2 & \mathrm{CNOT}_{C_{\mathrm r}\rightarrow M}\\
        3 & \mathrm{CNOT}_{M\rightarrow C_{\mathrm c}}\\
        4 & \mathrm{CNOT}_{C_{\mathrm r}\rightarrow M}
            \ \parallel\
            \mathrm{CNOT}_{C_{\mathrm c}\rightarrow B}.
    \end{array}
    \label{Seq:remote_CNOT}
\end{equation}
Here \(\parallel\) denotes simultaneous operations.
On computational-basis labels, the sequence acts as
\begin{equation}
    (a,0,m,0,b)
    \longmapsto
    (a,a,m,a,b\mathbin{\oplus}a),
    \label{Seq:remote_CNOT_action}
\end{equation}
where the entries are ordered as \((A,C_{\mathrm r},M,C_{\mathrm c},B)\).
Equation~\eqref{Seq:remote_CNOT_action} implements \(b\mapsto b\oplus a\) while restoring the input state of \(M\).
Measuring both buses in the \(X\) basis disentangles them from the atoms.
For outcomes \(s_{\mathrm r},s_{\mathrm c}\in\{0,1\}\), the remaining correction is \(Z_A^{s_{\mathrm r}\oplus s_{\mathrm c}}\), which can be tracked in the Pauli frame.
Thus, any atomic pair either shares a Cat Bus or can be connected by this constant-depth protocol, without atom transport.

This six-gate construction is optimal within the nearest-neighbour CNOT model on the path \(A-C_{\mathrm r}-M-C_{\mathrm c}-B\).
This statement assumes that the buses are initialized before, and measured after, the interaction.
Transferring the control from \(A\) to \(B\) requires at least one CNOT across each outer link.
The two links incident on \(M\) require at least four CNOTs in total.
If either link is used only once, the other must be used at least three times to transmit the control while restoring the independent state of \(M\).
Equation~\eqref{Seq:remote_CNOT} saturates the resulting lower bound of six cat--atom CNOTs.

\section{Dissipative error channel of the native
\texorpdfstring{$\mathrm{CZ}^{n}$}{CZn} gate}
\label{Sec:dissipative_error}

We analyse dissipation within the projected dynamics introduced in Section~\ref{Sec:gate_protocol}.
This treatment neglects corrections associated with the finite dissipative gap.
Dissipation acts during the common interaction interval $T_{\mathrm{int}}$; the preparation, return and virtual phase operations are treated as ideal.
We retain leading-order dissipative terms at zeroth order in the coherent-state overlap \(u\).
Section~\ref{Sec:crosstalk} treats finite-overlap corrections separately, whereas mixed finite-overlap--dissipative terms lie beyond the present approximation.
We first derive cat loss and Rydberg decay for one active target, then assemble the native-block channel.
We describe the open-system dynamics by the Lindblad master equation
\begin{align}
    \frac{\mathrm{d}\rho}{\mathrm{d}t}
    =\mathcal{L}\rho
    &=-i[\hat{H}_{\mathrm{CB}}^{(0)},\rho]
    +\mathscr{D}[\hat{L}_{1\mathrm{ph}}]\rho
    +\sum_{s\in\{r,r'\}}\mathscr{D}[\hat{L}_s]\rho,
\end{align}
where \(\hat{H}_{\mathrm{CB}}^{(0)}=g\alpha\,\sigma_x Z_{\mathrm{cat}}\) for one active target and
\begin{equation}
    \mathscr{D}[\hat{L}]\rho
    \equiv
    \hat{L}\rho\hat{L}^\dagger
    -\frac{1}{2}\{\hat{L}^\dagger\hat{L},\rho\}
\end{equation}
is the Lindblad dissipator.
The single-photon-loss jump operator is \(\hat{L}_{1\mathrm{ph}}=\sqrt{\kappa_1}\acat\).
For \(s\in\{r,r'\}\), Rydberg decay is described by
\begin{equation}
    \hat{L}_s=\sqrt{\gamma_s}\ket{\ell_s}\!\bra{s}.
\end{equation}
Here \(\ket{\ell_r}\) and \(\ket{\ell_{r'}}\) are effective leakage sinks that coarse-grain all decay products originating from \(\ket r\) and \(\ket{r'}\), respectively.
The sink labels include the associated environmental records and are taken to be orthogonal to each other and to the computational and Rydberg manifolds.

\subsection{Interaction picture and Dyson expansion}
In the interaction picture defined by
\(\hat{U}_{\mathrm{id}}(t)=e^{-i\hat{H}_{\mathrm{CB}}^{(0)}t}\), the density matrix
\(\tilde{\rho}(t)=\hat{U}_{\mathrm{id}}^\dagger(t)\rho(t)\hat{U}_{\mathrm{id}}(t)\) obeys
\begin{equation}
    \frac{\mathrm{d}\tilde{\rho}}{\mathrm{d}t}
    =\tilde{\mathcal{L}}(t)\tilde{\rho}(t)
    =\sum_k\mathscr{D}[\widetilde{L}_k(t)]\tilde{\rho},
\end{equation}
where
\(\widetilde{L}_k(t)=\hat{U}_{\mathrm{id}}^\dagger(t)\hat{L}_k\hat{U}_{\mathrm{id}}(t)\).
The error channel relative to the ideal interaction is the time-ordered Liouvillian propagator
\begin{align}
    \mathcal{E}_{\mathrm{error}}(\rho(0))
    &=\hat{U}^\dagger_{\mathrm{ideal}}\rho(T_{\mathrm{int}})\hat{U}_{\mathrm{ideal}}
    \\
    &=\mathcal{T}\exp\left(\int_0^{T_{\mathrm{int}}}\mathrm{d}t\,
    \tilde{\mathcal{L}}(t)\right)(\rho(0)),
\end{align}
where \(\hat{U}_{\mathrm{ideal}}=\hat{U}_{\mathrm{id}}(T_{\mathrm{int}})\).
For small integrated dissipative error probabilities, we retain the first-order Dyson expansion:
\begin{align}
    \mathcal{E}_{\mathrm{error}}
    &=\mathcal{I}
    +\delta\mathcal{E}_Z
    +\delta\mathcal{E}_{\mathrm{ryd}},
    \\
    \delta\mathcal{E}_Z
    &=\int_0^{T_{\mathrm{int}}}\mathrm{d}t\,
    \mathscr{D}[\widetilde{L}_{1\mathrm{ph}}(t)],
    \\
    \delta\mathcal{E}_{\mathrm{ryd}}
    &=\int_0^{T_{\mathrm{int}}}\mathrm{d}t\,
    \left(
    \mathscr{D}[\widetilde{L}_{r}(t)]
    +\mathscr{D}[\widetilde{L}_{r'}(t)]
    \right).
\end{align}
\subsection{Single-photon loss}
At zeroth order in \(u\), \(\acat=\alpha Z_{\mathrm{cat}}\) and
\([\hat{H}_{\mathrm{CB}}^{(0)},Z_{\mathrm{cat}}]=0\).
The interaction-picture jump operator is therefore constant:
\begin{equation}
    \widetilde{L}_{1\mathrm{ph}}(t)=\sqrt{\kappa_1}\alpha Z_{\mathrm{cat}}.
\end{equation}
Integration of the dissipator over $T_{\mathrm{int}} = \frac{\pi}{2g\alpha}$ gives
\begin{equation}
    \delta\mathcal{E}_{Z}
    =\int_0^{T_{\mathrm{int}}}\mathrm{d}t\,
    \kappa_1\alpha^2\mathscr{D}[Z_{\mathrm{cat}}]
    =p_{Z_{\mathrm{cat}}}\mathscr{D}[Z_{\mathrm{cat}}],
\end{equation}
where
\(p_{Z_{\mathrm{cat}}}=\kappa_1\alpha^2T_{\mathrm{int}}
=\pi\kappa_1\alpha/(2g)\).
The corresponding first-order phase-flip channel is
\begin{equation}
    \mathcal{Z}_{\mathrm{cat}}(p_{Z_{\mathrm{cat}}})[\rho]
    =
    (1-p_{Z_{\mathrm{cat}}})\rho
    +p_{Z_{\mathrm{cat}}}Z_{\mathrm{cat}}\rho Z_{\mathrm{cat}}
    +\mathcal O(p_{Z_{\mathrm{cat}}}^2).
\end{equation}

\subsection{Rydberg decay}
Let \(\theta(t)=g\alpha t\).
Under the ideal Hamiltonian dynamics, the Rydberg-decay jump operators become
\begin{align}
    \widetilde{L}_{r}(t)
    &=\sqrt{\gamma_r}\ket{\ell_r}
    \left[
    \cos\theta(t)\bra{r}
    -i\sin\theta(t)\bra{r'}Z_{\mathrm{cat}}
    \right],
    \\
    \widetilde{L}_{r'}(t)
    &=\sqrt{\gamma_{r'}}\ket{\ell_{r'}}
    \left[
    \cos\theta(t)\bra{r'}
    -i\sin\theta(t)\bra{r}Z_{\mathrm{cat}}
    \right].
\end{align}
Because \(\ket r\) and \(\ket{r'}\) have nearby principal quantum numbers, their radiative lifetimes are nearly equal.
We therefore set \(\gamma_r=\gamma_{r'}\equiv\gamma\).
The combined no-jump operator then reduces to the time-independent projector
\begin{equation}
    \widetilde{L}_r^\dagger\widetilde{L}_r
    +\widetilde{L}_{r'}^\dagger\widetilde{L}_{r'}
    =\gamma\left(\ket{r}\!\bra{r}+\ket{r'}\!\bra{r'}\right).
    \label{Seq:ryd_projector}
\end{equation}

Following the preparation pulse \(\hat V_1\), the atomic state is confined to
\(\operatorname{span}\{\ket0,\ket r\}\).
For the joint cat--atom state, write
\begin{equation}
    \rho(0)
    =
    \sum_{a,b\in\{0,r\}}
    \ket a\!\bra b\otimes\rho_{ab}^{(\mathrm{cat})},
    \qquad
    \rho_{ab}^{(\mathrm{cat})}
    \equiv\bra a\rho(0)\ket b,
\end{equation}
where each \(\rho_{ab}^{(\mathrm{cat})}\) is an operator on the Cat Bus.

Because the initial state contains no \(\ket{r'}\) component, the quantum-jump terms are
\begin{align}
    \widetilde{L}_r\rho(0)\widetilde{L}_r^\dagger
    &=
    \gamma\cos^2\theta(t)
    \ket{\ell_r}\!\bra{\ell_r}
    \otimes\rho_{rr}^{(\mathrm{cat})},
    \nonumber\\
    \widetilde{L}_{r'}\rho(0)\widetilde{L}_{r'}^\dagger
    &=
    \gamma\sin^2\theta(t)
    \ket{\ell_{r'}}\!\bra{\ell_{r'}}
    \otimes
    Z_{\mathrm{cat}}\rho_{rr}^{(\mathrm{cat})}Z_{\mathrm{cat}}.
\end{align}
Applying Eq.~\eqref{Seq:ryd_projector} gives
\begin{align}
    \left\{
    \sum_{s\in\{r,r'\}}
    \widetilde{L}_s^\dagger\widetilde{L}_s,
    \rho(0)
    \right\}
    &=2\gamma\ket r\!\bra r
    \otimes\rho_{rr}^{(\mathrm{cat})}
    \nonumber\\
    &\quad+\gamma
    \ket r\!\bra0\otimes\rho_{r0}^{(\mathrm{cat})}
    \nonumber\\
    &\quad+\gamma
    \ket0\!\bra r\otimes\rho_{0r}^{(\mathrm{cat})}.
\end{align}

Using
\(\int_0^{T_{\mathrm{int}}}\cos^2\theta(t)\,\mathrm dt
=\int_0^{T_{\mathrm{int}}}\sin^2\theta(t)\,\mathrm dt
=T_{\mathrm{int}}/2\), define
\begin{equation}
    p_{\mathrm{leak}}
    \equiv\frac{\gamma T_{\mathrm{int}}}{2}
    =\frac{\pi\gamma}{4g\alpha}.
\end{equation}
The first-order interaction-picture state is
\begin{align}
    \tilde{\rho}(T_{\mathrm{int}})
    &=\rho(0)
    +p_{\mathrm{leak}}
    \ket{\ell_r}\!\bra{\ell_r}
    \otimes\rho_{rr}^{(\mathrm{cat})}
    \nonumber\\
    &\quad
    +p_{\mathrm{leak}}
    \ket{\ell_{r'}}\!\bra{\ell_{r'}}
    \otimes
    Z_{\mathrm{cat}}\rho_{rr}^{(\mathrm{cat})}Z_{\mathrm{cat}}
    \nonumber\\
    &\quad
    -2p_{\mathrm{leak}}
    \ket r\!\bra r\otimes\rho_{rr}^{(\mathrm{cat})}
    \nonumber\\
    &\quad
    -p_{\mathrm{leak}}
    \ket r\!\bra0\otimes\rho_{r0}^{(\mathrm{cat})}
    \nonumber\\
    &\quad
    -p_{\mathrm{leak}}
    \ket0\!\bra r\otimes\rho_{0r}^{(\mathrm{cat})}
    \nonumber\\
    &\quad
    +\mathcal O(p_{\mathrm{leak}}^2).
\end{align}

Because the ideal interaction has already been removed in the error frame, the channel is returned to the computational input basis by
\(\hat V_1^\dagger=\hat V_1\), which maps \(\ket r\) to \(\ket1\).
For atom \(j\), define
\begin{equation}
    Z_j\equiv\ket0_j\!\bra0-\ket1_j\!\bra1,
    \qquad
    \rho_{11}^{(\mathrm{cat})}
    \equiv\bra1_j\rho\ket1_j.
\end{equation}
The resulting joint cat--atom Rydberg-decay channel is
\begin{align}
    \mathcal{E}_{\mathrm{ryd}}^{(c,j)}
    (p_{\mathrm{leak}})[\rho]
    &= (1-p_{\mathrm{leak}})\rho
    +\frac{p_{\mathrm{leak}}}{2}
    (Z_j\rho+\rho Z_j)
    \nonumber\\
    &\quad
    +p_{\mathrm{leak}}
    \ket{\ell_r}_j\!\bra{\ell_r}
    \otimes\rho_{11}^{(\mathrm{cat})}
    \nonumber\\
    &\quad
    +p_{\mathrm{leak}}
    \ket{\ell_{r'}}_j\!\bra{\ell_{r'}}
    \otimes
    Z_{\mathrm{cat}}\rho_{11}^{(\mathrm{cat})}Z_{\mathrm{cat}}
    \nonumber\\
    &\quad
    +\mathcal O(p_{\mathrm{leak}}^2).
    \label{Seq:joint_rydberg_decay_channel}
\end{align}
The two orthogonal sink branches together remove population from an initially occupied \(\ket1\) state with probability
\(2p_{\mathrm{leak}}=\gamma T_{\mathrm{int}}\).
Decay through the \(\ket{r'}\) branch additionally leaves a conditional
\(Z_{\mathrm{cat}}\) operation on the Cat Bus.

For completeness, in the ideal-interaction error frame, the process fidelity of a channel with Kraus operators \(\{K_\mu\}\) and a \(d\)-dimensional computational input space is
\begin{equation}
    F_{\mathrm{pro}}(\mathcal E)
    =\frac{1}{d^2}\sum_\mu
    \left|\Tr_{\mathrm{comp}}K_\mu\right|^2.
\end{equation}
For one cat--atom pair, \(d=4\), and the no-jump Kraus operator associated with Eq.~\eqref{Seq:joint_rydberg_decay_channel} is
\begin{equation}
    K_0
    =I_{\mathrm{cat}}\otimes
    \left[
    \ket0\!\bra0+(1-p_{\mathrm{leak}})\ket1\!\bra1
    \right]
    +\mathcal O(p_{\mathrm{leak}}^2).
\end{equation}
The leakage Kraus operators have zero trace on the computational input space, while
\(\Tr_{\mathrm{comp}}K_0=4-2p_{\mathrm{leak}}\).
Consequently,
\begin{equation}
    1-F_{\mathrm{pro}}
    =p_{\mathrm{leak}}+\mathcal O(p_{\mathrm{leak}}^2).
\end{equation}
Thus, although an occupied \(\ket1\) state loses population with probability \(2p_{\mathrm{leak}}\), the leading process infidelity of the cat--atom gate is \(p_{\mathrm{leak}}\).

\subsection{Native-block channel and fixed operating point}

Let $\mathcal{A}$ denote the set of atoms addressed during one native interaction block and let $n=|\mathcal{A}|$.
All addressed atoms share the same Cat Bus and the same interaction interval $T_{\mathrm{int}}$, independent of $n$.
The block-level cat-loss probability is
\begin{equation}
    p_{Z_{\mathrm{cat}}}
    =\frac{\pi\kappa_1\alpha}{2g}
\end{equation}
and the target-local Rydberg-decay probability is
\begin{equation}
    p_{\mathrm{leak}}
    =\frac{\pi\gamma}{4g\alpha}.
\end{equation}
To the order retained in the Dyson expansion, the dissipative channel of the native block is
\begin{equation}
    \begin{aligned}
        \mathcal{E}_{\mathrm{diss}}^{(\mathcal{A})}
        ={}&
        \mathcal{Z}_{\mathrm{cat}}(p_{Z_{\mathrm{cat}}})
        \circ
        \prod_{j\in\mathcal{A}}
        \mathcal{E}_{\mathrm{ryd}}^{(c,j)}(p_{\mathrm{leak}})\\
        &+\mathcal{O}\!\left(
        p_{Z_{\mathrm{cat}}}^2,
        p_{\mathrm{leak}}^2,
        p_{Z_{\mathrm{cat}}}p_{\mathrm{leak}}
        \right),
    \end{aligned}
    \label{Seq:native_block_dissipative_channel}
\end{equation}
where the cat phase-flip channel acts once on the entire block, whereas each Rydberg-decay channel acts jointly on the Cat Bus and one addressed atom.
The sink Kraus operators have zero trace on the computational input space, so each addressed atom contributes \(p_{\mathrm{leak}}\) to the leading process infidelity.
The corresponding leading process infidelity is
\begin{equation}
    \epsilon_n(\alpha)
    =p_{Z_{\mathrm{cat}}}+np_{\mathrm{leak}}
    =\frac{\pi}{2g}
    \left(
    \kappa_1\alpha+\frac{n\gamma}{2\alpha}
    \right).
    \label{Seq:native_block_infidelity}
\end{equation}

For a fixed fan-out, the exact minimum of Eq.~\eqref{Seq:native_block_infidelity} is
\begin{equation}
    \alpha_{\mathrm{opt}}^{(n)}
    =\sqrt{\frac{n\gamma}{2\kappa_1}},
    \qquad
    \epsilon_{n,\min}
    =\pi\sqrt{\frac{n}{2C}},
    \qquad
    C\equiv\frac{g^2}{\kappa_1\gamma}.
    \label{Seq:fanout_specific_optimum}
\end{equation}
At \(\gamma/\kappa_1=18\), this gives \(\alpha_{\mathrm{opt}}^{(1)}=3\) and \(\alpha_{\mathrm{opt}}^{(4)}=6\).
For the circuit-level comparison, however, \(p\) is defined as the process infidelity of the elementary two-qubit \(\mathrm{CZ}^{1}\) gate, matching the error parameter of the D2 model.
We therefore fix \(\alpha=3\) for both \(\mathrm{CZ}^{1}\) and \(\mathrm{CZ}^{4}\), rather than reoptimizing the cat amplitude for each fan-out.
At this operating point,
\begin{equation}
    p_{Z_{\mathrm{cat}}}=p_{\mathrm{leak}}=\frac{p}{2},
    \qquad
    p\equiv\epsilon_1.
    \label{Seq:single_target_reference_error}
\end{equation}
Their leading dissipative infidelities are consequently
\begin{equation}
    \epsilon_1=p,
    \qquad
    \epsilon_4=\frac{5}{2}p.
    \label{Seq:CZ1_CZ4_infidelities}
\end{equation}
Fan-out-specific optimization could reduce the intrinsic four-target error but is not included in this controlled benchmark.

\section{Finite-overlap corrections and multi-target crosstalk}
\label{Sec:crosstalk}

The order-$u$ term in Eq.~\eqref{Seq:Heff} breaks the leading-order factorization of the multi-target interaction.
We quantify the resulting coherent error after projecting the complete gate sequence onto the atomic computational subspace.
Here, \emph{finite-overlap correction} denotes the complete $u$-dependent error, including its single-target contribution.
We use \emph{multi-target crosstalk} only for the additional dependence on the other active targets.

\subsection{Computational-subspace error propagator and fidelity}

After the preparation pulse, each atom occupies either the dark state $\ket{0}$ or the active Rydberg state $\ket{r}$.
For atom $j$, define
\begin{align}
    P_{0,j}
    &=\ket{0}_j\!\bra{0},&
    Q_j
    &=\ket{r}_j\!\bra{r}+\ket{r'}_j\!\bra{r'}.
\end{align}
The Rydberg Pauli operators introduced in Section~\ref{Sec:gate_protocol} are partial Pauli operators on this three-level space and satisfy
\begin{equation}
    \sigma_{x,j}^2=\sigma_{y,j}^2=Q_j,
    \qquad
    \sigma_{x,j}P_{0,j}=\sigma_{y,j}P_{0,j}=0.
\end{equation}

The preparation, return and virtual phase operations are ideal and independent of \(u\).
They therefore cancel between the exact and ideal gate sequences in the relative error propagator.
It is therefore sufficient to work in the prepared basis after \(\hat V_1\) and compare the exact and ideal interaction evolutions.
At the ideal interaction time \(T_{\mathrm{int}}=\pi/(2g\alpha)\), the resulting error propagator is
\begin{align}
    \hat E(u)&=e^{i(\pi/2)\hat H_0}
    \exp\left[-i\frac{\pi}{2}
    \frac{\hat H_0+u\hat V}{\sqrt{1-u^2}}\right]\nonumber\\
    &=\exp[-i(u\hat K_1+u^2\hat K_2+\cdots)].
    \label{Seq:error_propagator_sigma}
\end{align}
Here, $\hat K_2$ denotes the complete Hermitian order-$u^2$ generator, including the Magnus commutator and the correction from the prefactor $1/\sqrt{1-u^2}$.
Using the dimensionless time $\tau=g\alpha t$, the first-order generator is
\begin{equation}
    \hat K_1=\int_0^{\pi/2}\mathrm{d}\tau\,
    e^{i\tau\hat H_0}\hat V e^{-i\tau\hat H_0}.
    \label{Seq:K1_sigma}
\end{equation}

Let $P_{\rm comp}$ project onto the prepared input space $\{\ket{0},\ket{r}\}^{\otimes n}$ and the complete cat qubit.
This space is unitarily equivalent, under the preparation pulse, to the original atomic computational space $\{\ket{0},\ket{1}\}^{\otimes n}$, and has dimension $d=2^{n+1}$.
Conjugation by $\hat H_0$ preserves the factor $\sigma_{y,j}$ in each contribution to Eq.~\eqref{Seq:K1_sigma}.
Consequently, every term in $\hat K_1$ either annihilates a dark atom or maps an active $\ket{r}_j$ to $\ket{r'}_j$.
The remaining factors act only on the other atoms.
It follows that
\begin{equation}
    P_{\rm comp}\hat K_1P_{\rm comp}=0.
    \label{Seq:PKP_zero_sigma}
\end{equation}
Thus, the order-$u$ correction produces finite-overlap-induced coherent leakage from the prepared computational subspace into the $\ket{r'}$ manifold.

Define the projected overlap operator
\begin{equation}
    \hat M(u)=P_{\rm comp}\hat E(u)P_{\rm comp}.
\end{equation}
Its Haar-averaged overlap with the ideal gate is~\cite{Pedersen2007}
\begin{equation}
    F_{\rm avg}
    =\frac{\Tr(\hat M\hat M^\dagger)+|\Tr\hat M|^2}{d(d+1)}.
    \label{Seq:Favg_sigma}
\end{equation}
For non-unitary $\hat M$, Eq.~\eqref{Seq:Favg_sigma} counts finite-overlap-induced coherent leakage from the computational output subspace as failure.

Expanding Eq.~\eqref{Seq:error_propagator_sigma} through second order and using Eq.~\eqref{Seq:PKP_zero_sigma} gives
\begin{align}
    \hat M={}&P_{\rm comp}
    -iu^2P_{\rm comp}\hat K_2P_{\rm comp}\nonumber\\
    &-\frac{u^2}{2}P_{\rm comp}\hat K_1^2P_{\rm comp}
    +\mathcal{O}(u^3).
\end{align}
The contribution from $\hat K_2$ cancels from Eq.~\eqref{Seq:Favg_sigma} at this order.
In addition, the operators defined in Eq.~\eqref{Seq:Heff} obey
\begin{align}
    \Zcat\hat H_0\Zcat&=\hat H_0,\nonumber\\
    \Zcat\hat V\Zcat&=-\hat V,\qquad
    [\Zcat,P_{\rm comp}]=0.
\end{align}
Hence $\hat M(-u)=\Zcat\hat M(u)\Zcat$ and $F_{\rm avg}(-u)=F_{\rm avg}(u)$.
The fidelity is an even function of $u$.
The leading infidelity is therefore
\begin{equation}
    1-F_{\rm avg}
    =\frac{u^2}{d}
    \Tr\!\left(P_{\rm comp}\hat K_1^2P_{\rm comp}\right)
    +\mathcal{O}(u^4).
    \label{Seq:fidelity_trace_sigma}
\end{equation}
The calculation thus reduces to an exact evaluation of the normalized trace of $\hat K_1^2$.

\subsection{First Magnus generator}

To evaluate Eq.~\eqref{Seq:K1_sigma}, introduce
\begin{equation}
    \hat B_j=\sum_{k\ne j}\sigma_{x,k}.
\end{equation}
The self term commutes, $[\sigma_{x,j}\Zcat,\sigma_{y,j}\Ycat]=0$, whereas the other active atoms rotate the shared cat operator.
The interaction-picture contribution associated with target $j$ is therefore
\begin{align}
    \hat V_j^{(I)}(\tau)
    &\equiv e^{i\tau\hat H_0}
    (\sigma_{y,j}\Ycat)e^{-i\tau\hat H_0}\nonumber\\
    &=\sigma_{y,j}
    \left[
    \Ycat\cos(2\tau\hat B_j)+\Xcat\sin(2\tau\hat B_j)
    \right].
    \label{Seq:interaction_picture_sigma}
\end{align}
Defining
\begin{align}
    \hat{\mathcal{J}}_j
    &=\int_0^\pi\mathrm{d}\theta\,e^{i\theta\hat B_j},\nonumber\\
    \hat{\mathcal{O}}_j&=
    \Ycat\frac{\hat{\mathcal{J}}_j+\hat{\mathcal{J}}_j^\dagger}{2}
    +\Xcat\frac{\hat{\mathcal{J}}_j-\hat{\mathcal{J}}_j^\dagger}{2i},
\end{align}
we obtain
\begin{equation}
    \hat K_1=-\frac12\sum_j\sigma_{y,j}\hat{\mathcal{O}}_j.
    \label{Seq:K1_sum_sigma}
\end{equation}

For a single target, $\hat B_1=0$ and $\hat{\mathcal{J}}_1=\pi I$, such that
\begin{equation}
    \hat K_1=-\frac{\pi}{2}\sigma_{y,1}\Ycat\neq0.
\end{equation}
Finite overlap therefore produces a single-target coherent-leakage amplitude.
For multiple targets, $\hat{\mathcal{O}}_j$ depends on the other active atoms through $\hat B_j$, and amplitudes associated with distinct targets can interfere.
This separates the single-target coherent-leakage amplitude from the dependence generated by other active targets.

\subsection{Evaluation in a fixed active-atom sector}

After the preparation pulse, a computational-basis state containing $q$ atoms in $\ket{1}$ contains exactly $q$ active Rydberg pseudospins in $\ket{r}$; the remaining $n-q$ atoms stay dark.
The trace in Eq.~\eqref{Seq:fidelity_trace_sigma} can therefore be decomposed into sectors of fixed active-atom number $q$.

The prepared computational space has the direct-sum structure
\begin{equation}
    \mathcal H_{\rm comp}
    =\bigoplus_{q=0}^{n}\mathcal H_q,
    \qquad
    \dim\mathcal H_q=2\binom nq,
    \label{Seq:active_sector_decomposition}
\end{equation}
where the factor of two accounts for the complete cat-qubit space.
Permutation symmetry makes the diagonal trace contribution identical for all atomic configurations in $\mathcal H_q$, so one representative configuration determines each sector.

For an active atom $j$, let $\nu=q-1$ denote the number of other active targets.
Within this sector,
\begin{equation}
    e^{i\theta\hat B_j}
    =\prod_{\substack{k=1\\k\ne j}}^{q}
    \left(\cos\theta\,I+i\sin\theta\,\sigma_{x,k}\right).
    \label{Seq:active_product_sigma}
\end{equation}
Define
\begin{equation}
    C_{\nu,m}=\int_0^\pi
    \cos^{\nu-m}\theta\,\sin^m\theta\,\mathrm{d}\theta.
    \label{Seq:CNm_sigma}
\end{equation}
Reflection about $\pi/2$ gives $C_{\nu,m}=0$ unless $m\equiv\nu\pmod 2$.
For the surviving terms,
\begin{equation}
    C_{\nu,m}
    =\frac{\Gamma[(\nu-m+1)/2]\Gamma[(m+1)/2]}
    {\Gamma(\nu/2+1)}.
    \label{Seq:CNm_gamma_sigma}
\end{equation}
For even $\nu$, the expansion includes the $m=0$ identity string.

Let $\ket{R_q}=\ket{r}^{\otimes q}$ denote the active part of a representative state in $\mathcal H_q$.
The inactive atoms remain in $\ket{0}$ and are annihilated by $\hat K_1$.
The cat-state-averaged coefficient of this sector is
\begin{equation}
    c_q=\frac12\Tr_{\rm cat}
    \bra{R_q}\hat K_1^2\ket{R_q}.
    \label{Seq:cq_def_sigma}
\end{equation}
A nonzero self contribution in $\hat K_1^2$ selects two identical neighbour strings.
Defining
\begin{equation}
    D_\nu=\sum_{m=0}^{\nu}\binom \nu m C_{\nu,m}^2,
    \label{Seq:DN_sum_sigma}
\end{equation}
the sum of the $q$ self terms is
\begin{equation}
    c_q^{\rm self}=\frac q4D_{q-1}.
\end{equation}

For an ordered pair $(j,k)$, a nonzero matrix element selects strings containing $\sigma_{x,k}$ for target $j$ and $\sigma_{x,j}$ for target $k$.
The choices on the remaining $m-1$ positions must match.
The multiplicity is $\binom{\nu-1}{m-1}$, while the atomic matrix element is
\begin{equation}
    \bra{r}\sigma_y\sigma_x\ket{r}
    \bra{r}\sigma_x\sigma_y\ket{r}
    =(-i)(i)=1.
\end{equation}
Thus, the ordered-pair contribution is non-negative.
Defining
\begin{equation}
    Q_\nu=\sum_{m=1}^{\nu}
    \binom{\nu-1}{m-1}C_{\nu,m}^2,\qquad \nu\ge1,
    \label{Seq:QN_sum_sigma}
\end{equation}
we obtain, for $q\ge2$,
\begin{equation}
    c_q^{\rm pair}=\frac{q(q-1)}4Q_{q-1},
    \qquad q\ge2,
\end{equation}
and hence
\begin{equation}
    \begin{aligned}
        c_0&=0,\qquad c_1=\frac{\pi^2}{4},\\
        c_q&=\frac q4D_{q-1}
        +\frac{q(q-1)}4Q_{q-1},\qquad q\ge2.
    \end{aligned}
    \label{Seq:cq_exact_sigma}
\end{equation}
The self term persists for a single target, whereas the ordered-pair term vanishes for $q=1$.
For $q>1$, the self term is also dressed by the other active atoms through $D_{q-1}$.

The sums entering Eq.~\eqref{Seq:cq_exact_sigma} have the exact integral representations
\begin{align}
    D_\nu&=\int_0^\pi\mathrm{d}\theta\int_0^\pi\mathrm{d}\phi\,
    \cos^\nu(\theta-\phi),\label{Seq:DN_integral_sigma}\\
    Q_\nu&=\int_0^\pi\mathrm{d}\theta\int_0^\pi\mathrm{d}\phi\,
    \sin\theta\sin\phi\,
    \cos^{\nu-1}(\theta-\phi).
    \label{Seq:QN_integral_sigma}
\end{align}

\subsection{Trace over the full computational space}

Using the decomposition in Eq.~\eqref{Seq:active_sector_decomposition} and the sector coefficient defined in Eq.~\eqref{Seq:cq_def_sigma}, the full trace is
\begin{equation}
    \Tr\!\left(P_{\rm comp}\hat K_1^2P_{\rm comp}\right)
    =2\sum_{q=0}^{n}\binom nq c_q.
    \label{Seq:trace_sector_sum_sigma}
\end{equation}
With $d=2^{n+1}$, the normalized coefficient is
\begin{equation}
    \frac{1}{d}
    \Tr\!\left(P_{\rm comp}\hat K_1^2P_{\rm comp}\right)
    =2^{-n}\sum_{q=0}^{n}\binom nq c_q
    \equiv\bar c_n.
    \label{Seq:cbar_sigma}
\end{equation}
Substitution into Eq.~\eqref{Seq:fidelity_trace_sigma} gives
\begin{equation}
    1-F_{\rm avg}=u^2\bar c_n+\mathcal{O}(u^4).
    \label{Seq:crosstalk_infidelity_sigma}
\end{equation}
For $n=1$, $\bar c_1=\pi^2/8$ gives the single-target finite-overlap contribution.

The finite-$n$ coefficient can be evaluated without approximation.
Using Eqs.~\eqref{Seq:DN_integral_sigma} and~\eqref{Seq:QN_integral_sigma}, define
\begin{equation}
    h(\theta,\phi)
    =\frac{1+\cos(\theta-\phi)}2
    =\cos^2\frac{\theta-\phi}{2}.
\end{equation}
The self contribution, and for $n\ge2$ the ordered-pair contribution, are then exactly
\begin{align}
    \bar c_n^{\rm self}
    &=\frac n8
    \int_0^\pi\mathrm{d}\theta\int_0^\pi\mathrm{d}\phi\,
    h(\theta,\phi)^{n-1},\nonumber\\
    \bar c_n^{\rm pair}
    &=\frac{n(n-1)}{16}
    \int_0^\pi\mathrm{d}\theta\int_0^\pi\mathrm{d}\phi\,
    \sin\theta\sin\phi\,
    h(\theta,\phi)^{n-2}.
    \label{Seq:cbar_integrals_sigma}
\end{align}
Eqs.~\eqref{Seq:cbar_sigma} and~\eqref{Seq:cbar_integrals_sigma} are exact for finite $n$, with $\bar c_1^{\rm pair}=0$.

\subsection{Large-fan-out asymptotics}

We now apply a single large-$n$ approximation to the exact trace in Eq.~\eqref{Seq:cbar_integrals_sigma}.
The kernel $h(\theta,\phi)$ is maximal along the one-dimensional ridge $\theta=\phi$.
Writing $\delta=\theta-\phi$, its local form is
\begin{equation}
    h(\theta,\phi)^s
    \simeq\exp\left[-\frac{s}{4}\delta^2\right],
    \qquad s\to\infty.
\end{equation}
Integration transverse to the ridge gives
\begin{align}
    \int_0^\pi\mathrm{d}\theta\int_0^\pi\mathrm{d}\phi\,h^s
    &\sim2\pi\sqrt{\frac{\pi}{s}},\nonumber\\
    \int_0^\pi\mathrm{d}\theta\int_0^\pi\mathrm{d}\phi\,
    \sin\theta\sin\phi\,h^s
    &\sim\pi\sqrt{\frac{\pi}{s}}.
\end{align}
Substitution into Eq.~\eqref{Seq:cbar_integrals_sigma} yields
\begin{align}
    \bar c_n^{\rm self}
    &\sim\frac{\pi\sqrt{\pi}}4n^{1/2},&
    \bar c_n^{\rm pair}
    &\sim\frac{\pi\sqrt{\pi}}{16}n^{3/2}.
\end{align}
The large-fan-out behaviour is therefore dominated by ordered pairs of active targets,
\begin{equation}
    \bar c_n\sim\frac{\pi\sqrt{\pi}}{16}n^{3/2},
    \qquad
    1-F_{\rm avg}
    \sim\frac{\pi\sqrt{\pi}}{16}
    n^{3/2}e^{-4|\alpha|^2}.
    \label{Seq:crosstalk_asymptotic_sigma}
\end{equation}

\subsection{Finite-overlap error for the
\texorpdfstring{$\mathrm{CZ}^{4}$}{CZ4} operation}

The Cat Bus scheduling considered in the main text implements a simultaneous four-target $\mathrm{CZ}^{4}$ operation, corresponding to a fan-out of $n=4$.
Evaluating the exact finite-$n$ coefficient in Eq.~\eqref{Seq:cbar_sigma} gives
\begin{equation}
    \bar c_4=\frac{\pi^2}{4}+\frac{19}{9}\simeq4.58.
\end{equation}
At the operating point $|\alpha|^2=9$,
\begin{equation}
    u^2=e^{-36}\simeq2.32\times10^{-16},
\end{equation}
and hence
\begin{align}
    1-F_{\rm avg}
    &=\left(\frac{\pi^2}{4}+\frac{19}{9}\right)e^{-36}
    +\mathcal{O}(e^{-72})\nonumber\\
    &\simeq1.06\times10^{-15}.
\end{align}

We independently verified Eq.~\eqref{Seq:cq_exact_sigma} using a collective-spin construction of \(\hat K_1\) for \(1\le q\le24\).
A full tensor-product construction provides a second check for \(1\le q\le6\).
Direct finite-\(u\) projected evolutions for \(n=1,2,4,8,12\) recover the coefficient in Eq.~\eqref{Seq:cbar_sigma} in the limit \(u\to0\).

\section{Cat-mediated syndrome extraction}
\label{Sec:syndrome_extraction}

\subsection{Minimum-vertex-cover-based star scheduling}

\begin{figure}[htbp]
    \centering
    \includegraphics[width=\linewidth]{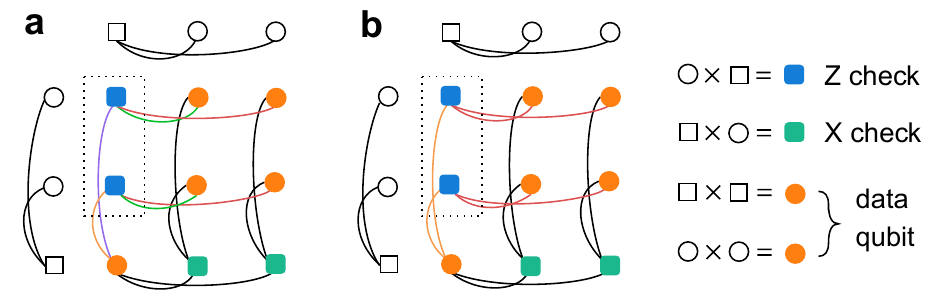}
    \caption{%
    \textbf{Comparison of matching- and MVC-based scheduling for the illustrated HGP-code instance.} Edge color denotes the scheduling step in both panels, such that all interactions with the same color are assigned to the same scheduling step.
    \textbf{a}, Matching-based scheduling requires four steps for the $Z$-check measurements.
    \textbf{b}, Minimum-vertex-cover-based star scheduling requires two steps by grouping interactions that share a common star center.
    }
    \label{fig:appendix_scheduling_coloration}
\end{figure}

Here we illustrate the effectiveness of the MVC-based star scheduling described in the main text by comparing it with conventional matching-based edge coloring. Conventional syndrome-extraction scheduling minimizes the depth of pairwise interactions by assigning disjoint interactions to the same scheduling step. This pairwise construction, however, does not fully exploit the native one-to-many, non-local interaction of the Cat Bus, which allows interactions sharing a common center to be executed together as a star block. The detailed MVC-based construction is given in Algorithms~1 and 2 of Methods.

Supplementary Fig.~\ref{fig:appendix_scheduling_coloration} compares the two approaches for a small HGP-code instance. In both panels, edge color denotes the scheduling step, such that all interactions with the same color are assigned to the same scheduling step. The matching-based edge coloring in panel~\textbf{a} requires four scheduling steps for the illustrated $Z$-check interactions. By grouping interactions incident on a common center into star blocks, the MVC-based construction in panel~\textbf{b} reduces the depth to two steps. This example demonstrates how star-based scheduling exploits the native multi-target capability of the Cat Bus. For larger or more general HGP codes, however, the number of selected MVC centers can increase with the size of the underlying Tanner graph, and the resulting scheduling depth can
scale less favorably than matching-based edge coloring. This reflects an inherent trade-off between exploiting native hardware-level parallelism and minimizing asymptotic scheduling depth for a given hardware topology.

\subsection{Detailed circuit implementation}

\begin{figure}[htbp]
    \centering
    \includegraphics[width=\linewidth]{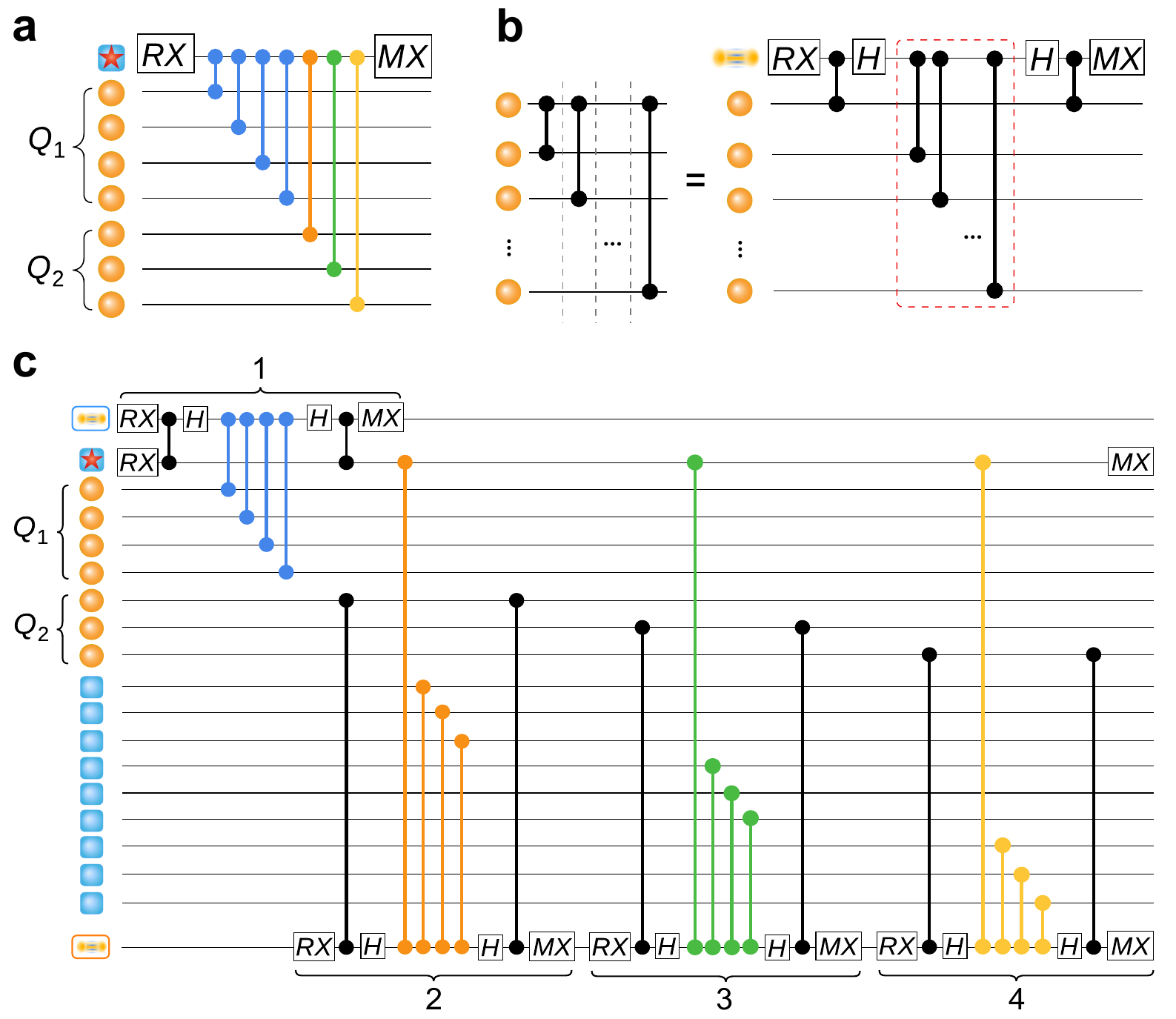}
    \caption{%
    \textbf{Circuit implementation of cat-mediated $Z$-check scheduling.}
    \textbf{a}, Syndrome-extraction circuit for the marked $Z$ check in Fig.~3.
    The MVC-based product scheduling partitions the circuit into four scheduling steps, indicated by different colors.
    \textbf{b}, Equivalent transformation implemented by the \texttt{ApplyCat} routine in Algorithm 3, which maps a multi-target atom-atom $\mathrm{CZ}^{n}$ operation to an equivalent circuit composed of native cat-atom $\mathrm{CZ}^{n}$ operations.
    The $\mathrm{CZ}^{n}$ block highlighted by a red dashed box is realized within one interaction interval.
    \textbf{c}, Complete cat-mediated circuit obtained by applying the transformation in \textbf{b} to the scheduling classes in \textbf{a}.
    }
    \label{sfig:appendix_circuit_weight7}
\end{figure}
Supplementary Fig.~\ref{sfig:appendix_circuit_weight7} translates the scheduling into an explicit syndrome-extraction circuit.
Note that in the circuit notation here, the $n$ same-colored $\mathrm{CZ}$ gates that share a common control and Cat Bus represent a single cat-mediated $\mathrm{CZ}^{n}$ block executed concurrently within one scheduling step, rather than $n$ independently scheduled pairwise $\mathrm{CZ}$ gates. The individual $\mathrm{CZ}$ gates are shown only to identify the participating targets.

Panel \textbf{a} shows the measurement circuit for the marked $Z$ check in Fig.~3 of the main text. The MVC-based product scheduling decomposes its $\mathrm{CZ}^{n}$ operations into four scheduling classes.

Panel \textbf{b} shows the equivalent transformation implemented by the \texttt{ApplyCat} routine in Methods Algorithm 3. This routine converts a multi-target $\mathrm{CZ}^{n}$ operation into a cat-mediated circuit of three cat--atom entangling layers. In the middle layer, highlighted by the red dashed box, one cat qubit couples concurrently to all $n$ target atoms, corresponding to a native cat-atom $\mathrm{CZ}^n$ gate.

Applying the \texttt{ApplyCat\_ZHorizontal} and \texttt{ApplyCat\_ZVertical} subroutines to each class gives the complete circuit in panel \textbf{c}.
During column-parallel $Z$-check measurements, atoms in the $Q_2$ region serve as control qubits.
Supplementary Fig.~\ref{sfig:appendix_circuit_weight7}\textbf{c} shows how the remaining $Z$ stabilizers are scheduled during parallel $\mathrm{CZ}^{4}$ operations.

\subsection{Cycle time comparison}

The syndrome-extraction cycle time is dominated by sequential layers of parallel $\mathrm{CZ}^{n}$ gates and by cat-qubit state preparation and measurement (SPAM).
For both $X$- and $Z$-check measurements, each cat-mediated atom--atom gate prepares the cat ancilla in $\ket{+}$ and concludes with an $X$-basis measurement.
Using $\kappa_2=2\pi\times2.16~\mathrm{MHz}$~\cite{Marquet2024Highkappa2} and $\alpha^2=9$, the estimates of Ref.~\cite{Christopher2022FTQC} give $T_{\mathrm{RX}}=10/(\kappa_2\alpha^2)\simeq82~\mathrm{ns}$ and $T_{\mathrm{MX}}=2/\kappa_2\simeq147~\mathrm{ns}$.
The SPAM duration entering the cycle-time estimate is therefore
\begin{equation}
    T_{\mathrm{SPAM}}
    =\frac{10}{\kappa_2\alpha^2}+\frac{2}{\kappa_2}
    \simeq229~\mathrm{ns}.
    \label{Seq:x_basis_spam_time}
\end{equation}
Together with $T_{\mathrm{gate}}=142~\mathrm{ns}$, this gives the non-pipelined Cat Bus cycle-time estimate
\begin{equation}
    T_{\mathrm{cycle}}^{\mathrm{cat}}
    =4n_{\mathrm C}\left(3T_{\mathrm{gate}}+T_{\mathrm{SPAM}}\right),
    \label{Seq:cat_bus_cycle_time}
\end{equation}
reported in Fig.~3d of the main text.

For comparison, the atom-rearrangement curve in Fig.~3d is evaluated using the transport model and experimental parameters of Xu \emph{et al.}~\cite{Xu2024ConstantOverhead}.
Let $L$ denote the number of atom-array sites along the dimension being rearranged.
The duration of one complete one-dimensional rearrangement layer is
\begin{equation}
    t_{\mathrm{rearr}}(L)
    =2\tau_{\mathrm t}\log_2L
    +(3+2\sqrt{2})
    \sqrt{\frac{6Ld_{\mathrm{lat}}}{a_{\mathrm p}}},
    \label{Seq:rearrangement_layer_time}
\end{equation}
where $\tau_{\mathrm t}$ is the transfer time between static spatial-light-modulator traps and dynamic acousto-optic-deflector traps, $a_{\mathrm p}$ is the peak transport acceleration and $d_{\mathrm{lat}}$ is the lattice spacing.
The non-pipelined product-coloration circuit contains $4\Delta_{\mathrm C}$ rearrangement layers per complete $X$- and $Z$-syndrome-extraction cycle, giving
\begin{equation}
    T_{\mathrm{cycle}}^{\mathrm{rearr}}
    =4\Delta_{\mathrm C}\,t_{\mathrm{rearr}}(L).
    \label{Seq:rearrangement_cycle_time}
\end{equation}
We use $\tau_{\mathrm t}=50~\mu\mathrm{s}$, $a_{\mathrm p}=0.02~\mu\mathrm{m}\,\mu\mathrm{s}^{-2}$ and $d_{\mathrm{lat}}=5~\mu\mathrm{m}$, following Ref.~\cite{Xu2024ConstantOverhead}.
The rearrangement benchmark retains the dominant transport time and neglects the shorter entangling-gate and readout durations.

The comparison in Fig.~3d uses non-pipelined schedules for both architectures.
For $r$ repeated syndrome-extraction cycles, pipelining reduces the Cat Bus scheduling depth from $4rn_{\mathrm C}$ to $(2r+2)n_{\mathrm C}$ and the product-coloration depth from $4r\Delta_{\mathrm C}$ to $(2r+2)\Delta_{\mathrm C}$~\cite{Xu2024ConstantOverhead}.
Both cycle times are therefore reduced by the same asymptotic factor of two, leaving the reported speed-up unchanged.

\section{Circuit-level simulations and decoding}
\label{Sec:simulations_decoding}
\subsection{Circuit-level noise models and decoder implementation}

We use a $\mathrm{CZ}$-limited circuit-level model to isolate entangling-gate noise and its propagation through the two syndrome-extraction architectures.
Single-qubit rotations, ancilla preparation and measurement, and idling are ideal in this controlled benchmark.
The retained noise therefore tests how the entangling-gate channel and circuit geometry determine logical performance.

For the hardware-derived model, let $b=(c_b,\mathcal{A}_b)$ denote a native interaction block, where $c_b$ is the Cat Bus and $\mathcal{A}_b$ is the set of simultaneously addressed atoms.
The simulations contain native $\mathrm{CZ}^{1}$ and $\mathrm{CZ}^{4}$ blocks, corresponding to $n_b=|\mathcal{A}_b|\in\{1,4\}$.
The simulator serializes the ideal operation as $\prod_{j\in\mathcal{A}_b}\mathrm{CZ}_{c_b,j}$.
Its hardware-derived noise, however, is assigned once to the native interaction block rather than independently to each pairwise gate.
The circuit simulator does not retain leakage levels or erasure information.
At the fixed operating point in Eq.~\eqref{Seq:single_target_reference_error}, we coarse-grain each joint Rydberg-decay channel in Eq.~\eqref{Seq:native_block_dissipative_channel}.
Its replacement is an unheralded single-qubit depolarizing channel with the same leading process infidelity.
The resulting hardware-derived stochastic block channel is
\begin{equation}
    \mathcal E_{\rm HD}^{(b)}(p)
    =
    \mathcal{Z}_{\mathrm{cat}}^{(c_b)}(p/2)
    \circ
    \prod_{j\in\mathcal{A}_b}\mathcal D_1^{(j)}(p/2),
    \label{Seq:hardware_stochastic_channel}
\end{equation}
Here $p=\epsilon_{\mathrm{CZ}^{1}}$ is the $\mathrm{CZ}^{1}$ process infidelity.
The channel $\mathcal{Z}_{\mathrm{cat}}^{(c_b)}(p/2)$ is sampled once per native block.
The channel $\mathcal D_1^{(j)}(p/2)$ is sampled independently for each addressed atom \(j\).
This coarse graining preserves the leading block infidelity but discards both the leakage-sink information and the conditional Cat Bus state.
Providing leakage or erasure information to the decoder would define a different noise model~\cite{Baranes2026Leveraging}.
The probability of at least one sampled fault in an $n_b$-target block is
\begin{equation}
    p_{\mathrm{block}}^{(n_b)}
    =
    1-\left(1-\frac{p}{2}\right)^{n_b+1}
    =
    \frac{n_b+1}{2}p+\mathcal O(p^2).
    \label{Seq:hardware_block_fault_probability}
\end{equation}
This gives $p_{\mathrm{block}}^{(1)}=p+\mathcal O(p^2)$ and $p_{\mathrm{block}}^{(4)}=5p/2+\mathcal O(p^2)$, consistent with Eq.~\eqref{Seq:CZ1_CZ4_infidelities}.
The comparison model is the standard two-qubit depolarizing channel $\mathcal D_2(p)$, which assigns probability $p/15$ to each non-identity two-qubit Pauli operator.

Following the space--time decoder of Xu \emph{et al.}~\cite{Xu2024ConstantOverhead}, we include a phenomenological depolarizing layer on all data qubits before the first noisy cycle.
This layer represents the residual data error passed from the preceding single-shot decoding block; it is a temporal-boundary prior rather than an additional physical gate.
Its probability value $p_{\rm r}=5p$ and the baseline BP iteration budget below are the HGP-code hyperparameters used in Ref.~\cite{Xu2024ConstantOverhead}.

We use non-overlapping $(3,3)$ sliding-window decoding~\cite{Huang2024Increasing,Kang2025quits}.
Each three-cycle space--time block is decoded by min-sum BP, after which its inferred update is committed.
The final noiseless data readout supplies the temporal boundary, after which we apply BP+ordered-statistics decoding using the OSD-e variant at order 15.
Thus OSD is used only at the final stage.
The maximum BP iteration counts are
\begin{equation}
 I_{\max}^{(0)}=N/5,
 \qquad I_{\max}^{(X)}=2.75\,N/5,
 \label{Seq:bp_iteration_budget}
\end{equation}
for syndrome-only and syndrome-plus-$X$-flag decoding, respectively.
Here $N$ is the number of data qubits.
The factor 2.75 compensates for the larger number of check equations because the $X$-flag detector count is 1.75 times the syndrome-detector count.
This choice approximately equalizes iterations per check equation, but not the wall-clock time.
The min-sum scaling factor is 0.625 for the controlled hardware-derived-versus-D2 noise-model comparison and 0.9 for the Cat Bus versus atom-rearrangement architecture comparison.

For a memory experiment containing $m$ syndrome-extraction cycles, let $P_{L}$ be the probability that at least one logical observable is flipped after final decoding.
We report the logical failure rate per cycle as
\begin{equation}
 \mathrm{LFR}=1-(1-P_L)^{1/m}.
 \label{Seq:lfr_definition}
\end{equation}
For Fig.~4\textbf{a,b} of the main text, the Monte Carlo stopping rule depends on the estimated per-cycle LFR.
We use $(N_{\max},E_{\max})=(4\times10^4,2500)$ in the low-LFR regime and $(10^4,10^4)$ when the estimated LFR exceeds 0.1.
Here $N_{\max}$ and $E_{\max}$ denote the maximum numbers of samples and observed failures, respectively.

We define the hardware-derived circuit-level threshold using the $\mathrm{CZ}^{1}$ process infidelity $p$ in Eq.~\eqref{Seq:single_target_reference_error}.
It is the critical value below which the per-cycle LFR decreases with increasing code distance.
The subthreshold fits below use the Monte Carlo estimates and standard errors reported for the threshold comparisons in Fig.~4 of the main text.

\subsection{Finite-cycle effects in threshold evaluation}

\begin{figure}[htbp]
    \centering
    \includegraphics[width=\linewidth]{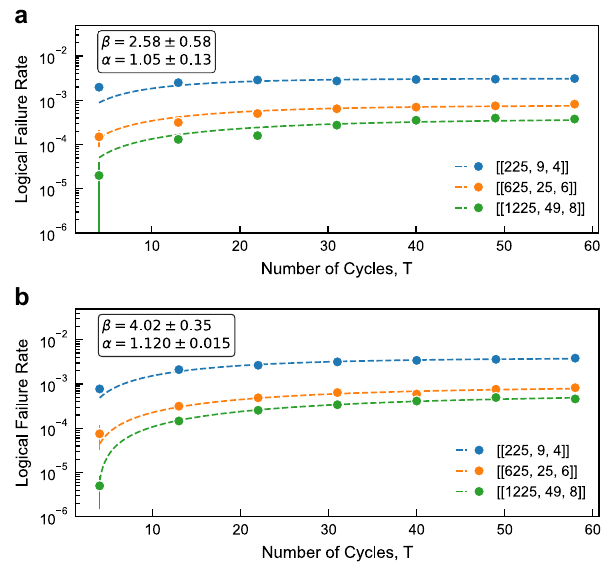}
    \caption{%
    \textbf{Finite-cycle relaxation of the logical failure rate.}
    Symbols show 10,000-sample Monte Carlo estimates at a physical error rate of $0.15\%$; dashed curves show fits to Eq.~\eqref{Seq:LFR_relaxation_process}.
    Both panels use the Cat Bus architecture with $X$-flag-assisted decoding.
    \textbf{a}, D2 error model.
    \textbf{b}, Hardware-derived error model.
    }
    \label{sfig:nclist_hardware-derived_Xflag}
\end{figure}
Finite-cycle threshold estimates require code distances to be compared at the same stage of temporal relaxation.
We therefore simulate the per-cycle LFR as a function of the number of syndrome-extraction cycles $T$ at fixed physical error rate $p=0.15\%$.
For both noise models in Supplementary Fig.~\ref{sfig:nclist_hardware-derived_Xflag}, the relaxation is described by
\begin{equation}
    \label{Seq:LFR_relaxation_process}
    \mathrm{LFR}(p,d,T)=\mathrm{LFR}(p,d,\infty) [1-\alpha \exp(-\frac{T}{\beta \cdot d})].
\end{equation}
The fitted parameters for $X$-flag-assisted decoding are
\begin{align}
    \alpha_{\rm D2}&=1.05\pm0.13,
    &\beta_{\rm D2}&=2.58\pm0.58, \notag\\
    \alpha_{\rm HD}&=1.120\pm0.015,
    &\beta_{\rm HD}&=4.02\pm0.35.
    \label{Seq:LFR_relaxation_fits}
\end{align}
The relaxation time therefore scales as $\beta d$, rather than remaining constant with code distance.
Companion fits without flag information show only a weak change in $\beta$ for a fixed architecture and noise model.
Changing the noise model produces a larger shift, and atom rearrangement gives a smaller $\beta$ than the Cat Bus under D2 noise.

For each controlled comparison in the main text, we consequently choose $T/d$ from the largest relaxation scale of the two datasets being compared.
The Cat Bus--atom-rearrangement comparison under D2 noise uses $T=3d$, whereas the hardware-derived--D2 comparison on the Cat Bus uses $T=4.5d$.
In both cases, the same $T/d$ is applied to the two datasets, with $T\simeq1.1\beta_{\max}d$.
This distance-dependent cycle count compares the codes at equivalent stages of temporal relaxation~\cite{Huang2024Increasing,Kang2025quits,Xu2024ConstantOverhead}.


\subsection{Decoding with cat-flag information}

The cat ancillas are measured before reset at the end of each syndrome-extraction cycle.
Their outcomes provide detector events in addition to the stabilizer-syndrome differences.
We call these events \emph{cat flags}, following the use of ancilla measurements to expose propagated faults in flag-based syndrome extraction~\cite{Chao2018Flag,ChamberlandBeverland2018Flag}.
We next identify which cat flags inform the logical Pauli-frame update used by the sliding-window decoder.

We construct one detector-error model (DEM) containing all syndrome and flag detectors, then project its detector rows.
This procedure holds the fault mechanisms and their probabilities fixed across comparisons.
If $\bm c$ is the binary mechanism vector of this reference DEM, the three nested observation records are
\begin{equation}
    Y_0=D,\qquad
    Y_X=(D,F_X),\qquad
    Y_{\rm all}=(D,F_X,F_Z),
    \label{Seq:flag_observations}
\end{equation}
where $D$ denotes the ordinary detection events and $F_X$ and $F_Z$ the two sets of cat-flag events.
For the first three-round decoding window, the target variable is the local logical Pauli-frame increment
\begin{equation}
    \bm\Lambda=L_1\bm c\pmod 2 .
    \label{Seq:local_logical_frame}
\end{equation}
This quantity is distinct from the boundary state passed to the next window and from the final logical observable of the complete memory experiment.

For an estimator $\widehat{\bm\Lambda}(Y)$ constructed from the available detector record, define the maximum-likelihood (ML) decision error as
\begin{equation}
 P_{\rm err}^{\rm ML}(\bm\Lambda;Y,p)
 \equiv
 \min_{\widehat{\bm\Lambda}}
 \Pr_p\!\left[\widehat{\bm\Lambda}(Y)\ne\bm\Lambda\right].
 \label{Seq:flag_ml_error}
\end{equation}
Thus, $P_{\rm err}^{\rm ML}$ is the smallest probability of assigning an incorrect local logical Pauli-frame increment using only the information in $Y$.
Let $w_{y,\lambda}$ be the coefficient of $p$ in the probability of the joint event $(Y=y,\bm\Lambda=\lambda)$.
Since the no-fault configuration dominates the record $y=0$, the low-error expansion is
\begin{equation}
 P_{\rm err}^{\rm ML}(\bm\Lambda;Y,p)
 =c_{1,\Lambda}^{\rm ML}(Y)p+\mathcal{O}(p^2),
\end{equation}
with
\begin{equation}
 c_{1,\Lambda}^{\rm ML}(Y)
 =
 \sum_{\lambda\ne0}w_{0,\lambda}
 +
 \sum_{y\ne0}
 \left(
 \sum_\lambda w_{y,\lambda}-\max_\lambda w_{y,\lambda}
 \right).
 \label{Seq:flag_ml_coefficient}
\end{equation}
Eq.~\eqref{Seq:flag_ml_coefficient} measures the irreducible first-order uncertainty in the logical-frame update for the specified detector record \cite{Poulin2006ML}.
The DEMs and their detector projections are generated with Stim \cite{Gidney2021Stim}.

The records in Eq.~\eqref{Seq:flag_observations} form a nested refinement: $Y_0$ is obtained from $Y_X$ by discarding $F_X$, and $Y_X$ is obtained from $Y_{\rm all}$ by discarding $F_Z$.
Any decision rule available for a coarse record can therefore be implemented for a refined record by ignoring the additional flags.
It follows that
\begin{align}
 P_{\rm err}^{\rm ML}(\bm\Lambda;Y_{\rm all},p)
 &\le P_{\rm err}^{\rm ML}(\bm\Lambda;Y_X,p)
 \le P_{\rm err}^{\rm ML}(\bm\Lambda;Y_0,p), \notag\\
 c_{1,\Lambda}^{\rm ML}(Y_{\rm all})
 &\le c_{1,\Lambda}^{\rm ML}(Y_X)
 \le c_{1,\Lambda}^{\rm ML}(Y_0).
 \label{Seq:flag_information_monotonicity}
\end{align}
Equivalently, splitting a coarse record class by a flag cannot reduce the summed maximum logical-class weights of the refined classes.
Additional flag information can therefore leave the optimal decision error unchanged or reduce it, but cannot increase it.

The two noise models differ in their first-order fault alphabets.
For a native block $b=(c_b,\mathcal A_b)$, the hardware-derived model contains one block-level cat fault and target-local atomic faults,
\begin{equation}
    \mathcal A_{\rm HD}^{(1)}(b)
    =
    \{Z_{\mathrm{cat}}^{(c_b)}\}
    \cup
    \bigcup_{j\in\mathcal A_b}
    \{X_j,Y_j,Z_j\},
    \label{Seq:hardware_fault_alphabet}
\end{equation}
whereas the D2 model assigns first-order probability to all 15 non-identity two-qubit Paulis following each $\mathrm{CZ}$ gate.
In particular, D2 contains nine double-sided components $P_{\rm d}P_{\rm c}$ with $P_{\rm d},P_{\rm c}\ne I$.
Some components produce the same syndrome record $D$ but different values of $\bm\Lambda$.
The correlated cat component is then resolved by $F_X$.
These double-sided events enter the hardware-derived model only at order $p^2$, so $F_X$ does not refine its leading-order logical-frame decision.

Supplementary Table~\ref{tab:flag_distinguishability} gives the resulting coefficients for the $N=225$ HGP code.
The analysis uses the logical-$Z$ sector of the first three-round window, which is the sector relevant to the $X$-memory simulation.
\begin{table}[htbp]
    \centering
    \caption{%
    Leading maximum-likelihood decision coefficient $c_{1,\Lambda}^{\rm ML}$ for the local logical Pauli-frame increment in the first three-round decoding window of the $N=225$ HGP code.
    }
    \label{tab:flag_distinguishability}
    \small
    \setlength{\tabcolsep}{6pt}
    \begin{tabular}{lcc}
        \hline \hline
        Detector record
        & Hardware-derived
        & D2 \\
        \hline
        $D$                 & 15.00 & 66.00 \\
        $(D,F_X)$           & 15.00 & 60.00 \\
        $(D,F_X,F_Z)$       & 15.00 & 60.00 \\
        \hline \hline
    \end{tabular}
\end{table}

For the hardware-derived channel, the coefficient is unchanged, $15.00\rightarrow15.00$.
For D2 noise, adding $F_X$ reduces it from $66.00$ to $60.00$, a relative reduction of $9.09\%$.
The reduction is distributed over 33 ordinary-record classes.
Adding $F_Z$ produces no further first-order reduction.
The same pattern occurs for $N=625$.
The hardware-derived coefficient remains $41.67$ for all three records.
For D2 noise, $F_X$ reduces the coefficient from $183.33$ to $166.67$, whereas $F_Z$ produces no further change.

The monotonicity in Eq.~\eqref{Seq:flag_information_monotonicity} applies to optimal inference on a fixed DEM ensemble.
It does not guarantee improved performance for finite BP+OSD.
The approximate decoder also depends on finite iteration and OSD budgets, factor-graph loops and committed sliding-window boundaries.
Enlarging the detector graph can therefore alter convergence even when the optimal risk cannot increase.
We omit cat flags from the controlled hardware-derived--D2 comparison, in which the circuit and detector set are held fixed.
For the D2 architecture comparison, we include $F_X$ because it provides first-order logical-frame information absent from $D$.
We omit $F_Z$ because it provides no additional first-order reduction.

\subsection{Threshold and subthreshold fitting}

We define each subthreshold fitting window independently of the fit.
For each controlled comparison, we retain sampled physical error rates satisfying
\begin{equation}
    0<p\leq 0.9p_{\mathrm{th}},
    \label{Seq:subthreshold_window}
\end{equation}
where $p_{\mathrm{th}}$ is obtained independently from the corresponding threshold-crossing analysis.
Because $p$ is sampled on a discrete grid, the largest retained value $p_{\max}$ can lie below $0.9p_{\mathrm{th}}$.
For example, the 4.5$d$ D2 dataset has $p_{\mathrm{th}}=0.465\%$, so the largest retained point is $p_{\max}=0.4167\%\simeq0.896p_{\mathrm{th}}$.

We fit each dataset globally across the $N=225$, 625 and 1225 code instances using the empirical ansatz
\begin{equation}
    \mathrm{LFR}(p,N)
    =A\left(\frac{p}{p_0}\right)^{\alpha N^{\beta}}.
    \label{Seq:subthreshold_scaling_ansatz}
\end{equation}
Here $p_0$ is a fitted reference error rate, not the independently estimated threshold.
We fit $\log_{10}(\mathrm{LFR})$ and propagate an LFR standard error $\delta\mathrm{LFR}$ as
\begin{equation}
    \sigma_{\log_{10}\mathrm{LFR}}
    =\frac{\delta\mathrm{LFR}}{\mathrm{LFR}\ln 10},
    \label{Seq:subthreshold_log_uncertainty}
\end{equation}
The nonlinear least-squares fit uses $\sigma_{\log_{10}\mathrm{LFR}}^{-2}$ as its weight.

\begin{figure*}[t]
    \centering
    \includegraphics[width=\textwidth]{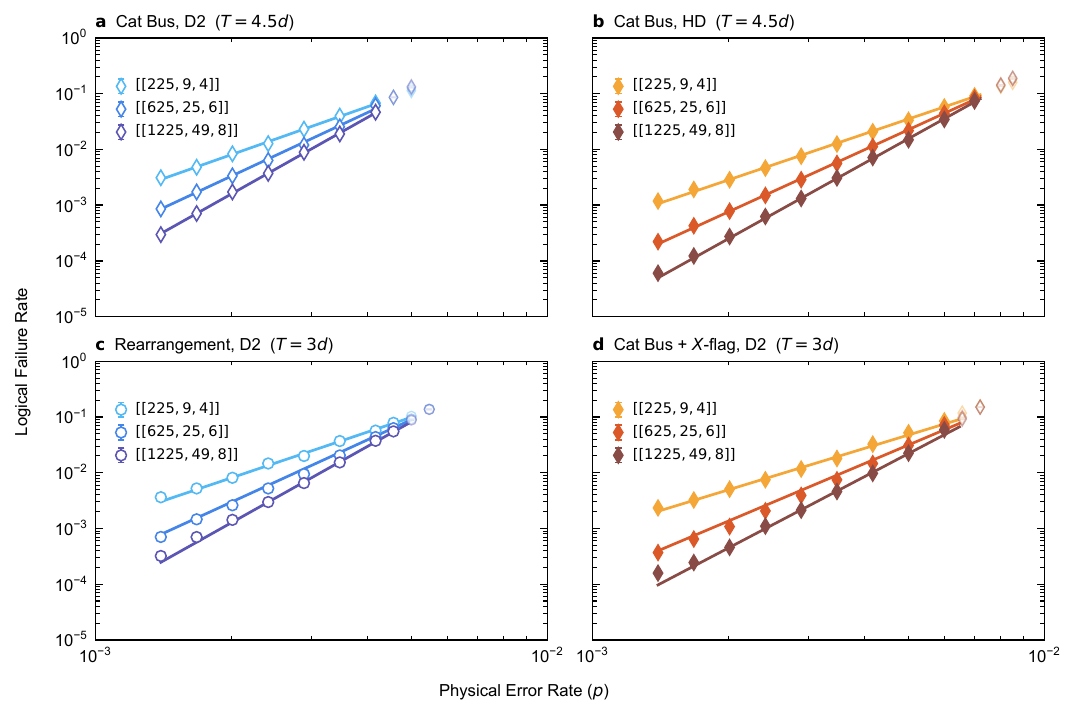}
    \caption{%
    \textbf{Weighted subthreshold scaling fits.}
    \textbf{a}, Cat Bus under D2 noise with $T=4.5d$ syndrome-extraction cycles.
    \textbf{b}, Cat Bus under the hardware-derived (HD) noise model with $T=4.5d$.
    \textbf{c}, Atom rearrangement under D2 noise with $T=3d$.
    \textbf{d}, Cat Bus with $X$-flag-assisted decoding under D2 noise with $T=3d$.
    Symbols show Monte Carlo estimates for the $[[225,9,4]]$, $[[625,25,6]]$ and $[[1225,49,8]]$ codes.
    Error bars denote one standard error propagated from binomial sampling and are smaller than the markers where not visible.
    Solid curves show the weighted global fits to Eq.~\eqref{Seq:subthreshold_scaling_ansatz} and extend to $0.9p_{\mathrm{th}}$.
    Faded open symbols lie outside the fitting windows and are excluded from the fits.
    }
    \label{Sfig:subthreshold_fits}
\end{figure*}

Supplementary Fig.~\ref{Sfig:subthreshold_fits} shows the twelve size-resolved curves, and Supplementary Table~\ref{Stab:subthreshold_fit_parameters} reports the four global fits.
Across the four controlled settings, the fitted exponent $\beta$ ranges from 0.276 to 0.320, corresponding to a stretched-exponential suppression with system size at fixed $p<p_0$.
The reduced chi-squared values exceed unity for three datasets, revealing systematic residuals beyond the Monte Carlo standard errors.
Equation~\eqref{Seq:subthreshold_scaling_ansatz} should therefore be interpreted as an empirical description of the sampled window, not as an exact asymptotic law.

\begin{table*}[t]
    \centering
    \caption{\textbf{Weighted subthreshold scaling fits.}
    Each dataset is fitted to $\mathrm{LFR}=A(p/p_0)^{\alpha N^\beta}$ using sampled points with $p\leq0.9p_{\mathrm{th}}$.
    The threshold $p_{\mathrm{th}}$ is obtained independently, and $p_{\max}$ is the largest sampled value retained.
    The column $n_p$ gives the number of physical-error values for each code instance ($N=225$, 625 and 1225).
    Fits use weighted nonlinear least squares in $\log_{10}(\mathrm{LFR})$ with propagated Monte Carlo standard errors.
    Parameter uncertainties are one standard deviation from the fit covariance; $\chi^2_\nu$ is the reduced chi-squared.}
    \label{Stab:subthreshold_fit_parameters}
    \small
    \setlength{\tabcolsep}{4.0pt}
    \renewcommand{\arraystretch}{1.15}
    \begin{tabular}{lcccccccc}
        \hline\hline
        Dataset & $p_{\mathrm{th}}$ (\%) & $p_{\max}$ (\%) & $n_p$ & $A$ & $p_0$ (\%) & $\alpha$ & $\beta$ & $\chi^2_\nu$ \\
        \hline
        Cat Bus, D2 & 0.465 & 0.417 & 7 & 0.122 $\pm$ 0.017 & 0.518 $\pm$ 0.020 & 0.639 $\pm$ 0.055 & 0.276 $\pm$ 0.013 & 1.29 \\
        Cat Bus, HD & 0.800 & 0.700 & 10 & 0.124 $\pm$ 0.004 & 0.792 $\pm$ 0.006 & 0.562 $\pm$ 0.013 & 0.293 $\pm$ 0.004 & 4.79 \\
        Rearrangement, D2 & 0.548 & 0.456 & 8 & 0.131 $\pm$ 0.006 & 0.554 $\pm$ 0.007 & 0.533 $\pm$ 0.028 & 0.301 $\pm$ 0.008 & 4.62 \\
        Cat Bus + $X$-flag, D2 & 0.720 & 0.600 & 9 & 0.138 $\pm$ 0.012 & 0.770 $\pm$ 0.020 & 0.436 $\pm$ 0.023 & 0.320 $\pm$ 0.008 & 8.59 \\
        \hline\hline
    \end{tabular}
\end{table*}

\clearpage

\renewcommand{\refname}{Supplementary References}
\bibliographystyle{apsrev4-2}
\bibliography{references}